\documentclass[aps,physrev,twocolumn,groupedaddress,superscriptaddress]{revtex4-2}
\usepackage{amssymb,mathrsfs}
\usepackage{placeins}
\usepackage{bm}
\usepackage{amsmath}
\usepackage{enumerate,algorithmicx,algorithm}
\usepackage{algpseudocode}

\usepackage{graphicx}
\usepackage{xcolor}
\DeclareMathOperator{\erf}{erf}

\usepackage{xr}
\usepackage[utf8]{inputenc}
\usepackage[T1]{fontenc}
\usepackage{array,multirow}
\usepackage{hyperref}
\usepackage{flushend}
\usepackage{float}

\makeatletter
\newcommand*{\addFileDependency}[1]{
  \typeout{(#1)}
  \@addtofilelist{#1}
  \IfFileExists{#1}{}{\typeout{No file #1.}}
}
\makeatother

\usepackage{float}
\begin{document}


\title{Machine-Learning Interatomic Potentials for Long-Range Systems} 


\author{Yajie Ji}
\email{jiyajie595@sjtu.edu.cn}
\affiliation{School of Mathematical Sciences, Shanghai Jiao Tong University, Shanghai 200240, China}
\author{Jiuyang Liang}
\email[Contact author: ]{jliang@flatironinstitute.org}
\affiliation{School of Mathematical Sciences, Shanghai Jiao Tong University, Shanghai 200240, China}
\affiliation{
Center for Computational Mathematics, Flatiron Institute, Simons Foundation, New York 10010, USA}
\author{Zhenli Xu}
\email[Contact author: ]{xuzl@sjtu.edu.cn}
\affiliation{School of Mathematical Sciences, Shanghai Jiao Tong University, Shanghai 200240, China}
\affiliation{CMA-Shanghai and MOE-LSC, Shanghai Jiao Tong University, Shanghai 200240, China}






\begin{abstract}
Machine-learning interatomic potentials have emerged as a revolutionary class of force-field models in molecular simulations, delivering quantum-mechanical accuracy at a fraction of the computational cost and enabling the simulation of large-scale systems over extended timescales. However, they often focus on modeling local environments, neglecting crucial long-range interactions. 
We propose a Sum-of-Gaussians Neural Network (SOG-Net), a lightweight and versatile framework for integrating long-range interactions into machine learning force field. The SOG-Net employs a latent-variable learning network that seamlessly bridges short-range and long-range components, coupled with an efficient Fourier convolution layer that incorporates long-range effects. By learning sum-of-Gaussians multipliers across different convolution layers, the SOG-Net adaptively captures diverse long-range decay behaviors while maintaining close-to-linear computational complexity during training and simulation via non-uniform fast Fourier transforms. The method is demonstrated effective for a broad range of long-range systems. 
\end{abstract}


\maketitle

Machine-learning interatomic potentials (MLIPs) have become essential tools in atomic force field modeling, bridging the gap between the high accuracy of computationally expensive quantum mechanical calculations and the efficiency of empirical force fields~\cite{friederich2021machine,unke2021machine,zhang2018deep}. Most MLIPs assume that the potential energy surface (PES) of an atom is determined by its local environment within a finite cutoff radius, an assumption rooted in the ``nearsightedness principle'' of electronic structure theory~\cite{kohn1996density,PNAS2005}. This approximation simplifies the model, enabling predictions to scale linearly with system size. However, it systematically neglects long-range (LR) interactions such as Coulomb, dispersion, and dipole forces, which are crucial for accurately modeling the heterogeneous structures and functionalities of polar materials and biological systems in molecular dynamics (MD) simulations~\cite{zhou2018electrostatic,PhysRevLett.132.228101}. Incorporating LR interactions into MLIP models while preserving efficiency and scalability remains a significant challenge in the field.

To capture LR effects in MLIP models, most efforts focused on electrostatic interactions. Early approaches~\cite{bartok2010gaussian,Unke2021} assigned fixed point charges with empirical force-field corrections; however, these methods struggled with charge-transfer effects and the challenge of defining accurate corrections. Recent advancements include the use of virtual charge sites, as employed in deep potential LR~\cite{zhang2022deep} and self-consistent field neural networks~\cite{gao2022self} as well as methods that predict effective partial charges while enforcing global charge conservation~\cite{unke2019physnet,ko2021fourth,shaidu2024incorporating}. In these models, once latent variables such as virtual sites or partial charges are determined, electrostatic interactions are computed using classical algorithms such as Ewald summation~\cite{ewald1921berechnung} or self-consistent field iterations. These approaches have demonstrated success in applications such as phase diagram predictions~\cite{zhang2021phase,bore2023realistic}. 
However, accurately and efficiently modeling diverse LR decay behaviors in MLIPs remains far from settled. The classical Ewald summation~\cite{ewald1921berechnung}, commonly used to account for LR effects, was originally designed for $1/r$ Coulomb potentials and is ill-suited for other LR decay rates, such as $1/r^p$ (with $p>1$ or non-integer), $e^{-\mu r}/r$ (with $\mu>0$), or more complex forms involving combinations of decay exponents~\cite{israelachvili2011intermolecular}.
Approaches such as the long-distance equivariant (LODE) method~\cite{huguenin2023physics,Grisafi2019JCP} improve the accuracy for inverse power-law decays but rely heavily on the accuracy of the initial LR tail guess. Similarly, message passing neural networks~\cite{schutt2017schnet,batzner20223,batatia2022mace,kosmala2023ewald,deng2023chgnet} capture nonlocal interactions through stacked graph convolution layers but involve prohibitive computational cost for slowly decaying LR potentials.

This paper proposes a sum-of-Gaussians neural network (SOG-Net) for accurately learning interatomic potentials with general LR tails. It separates atomic interactions into short-range (SR) and LR components, which are connected through a network that learns latent variables. These variables are processed by an efficient LR convolution layer that accounts for LR contributions, where the SOG multiplier in each layer is represented as a sum of Gaussians with trainable variances and amplitudes. The resulting SOG-Net effectively captures LR effects while maintaining nearly linear computational complexity during both training and simulation through the use of non-uniform fast Fourier transform (NUFFT)~\cite{dutt1993fast,greengard2004accelerating}. Unlike existing works relying on either various forms of Ewald summation or self-consistent iterations to account for LR contributions, our method adaptively captures diverse LR decay tails by learning Gaussian multipliers across different convolution layers. 

\begin{figure}[!ht] 
    \centering    \includegraphics[width=0.48\textwidth]{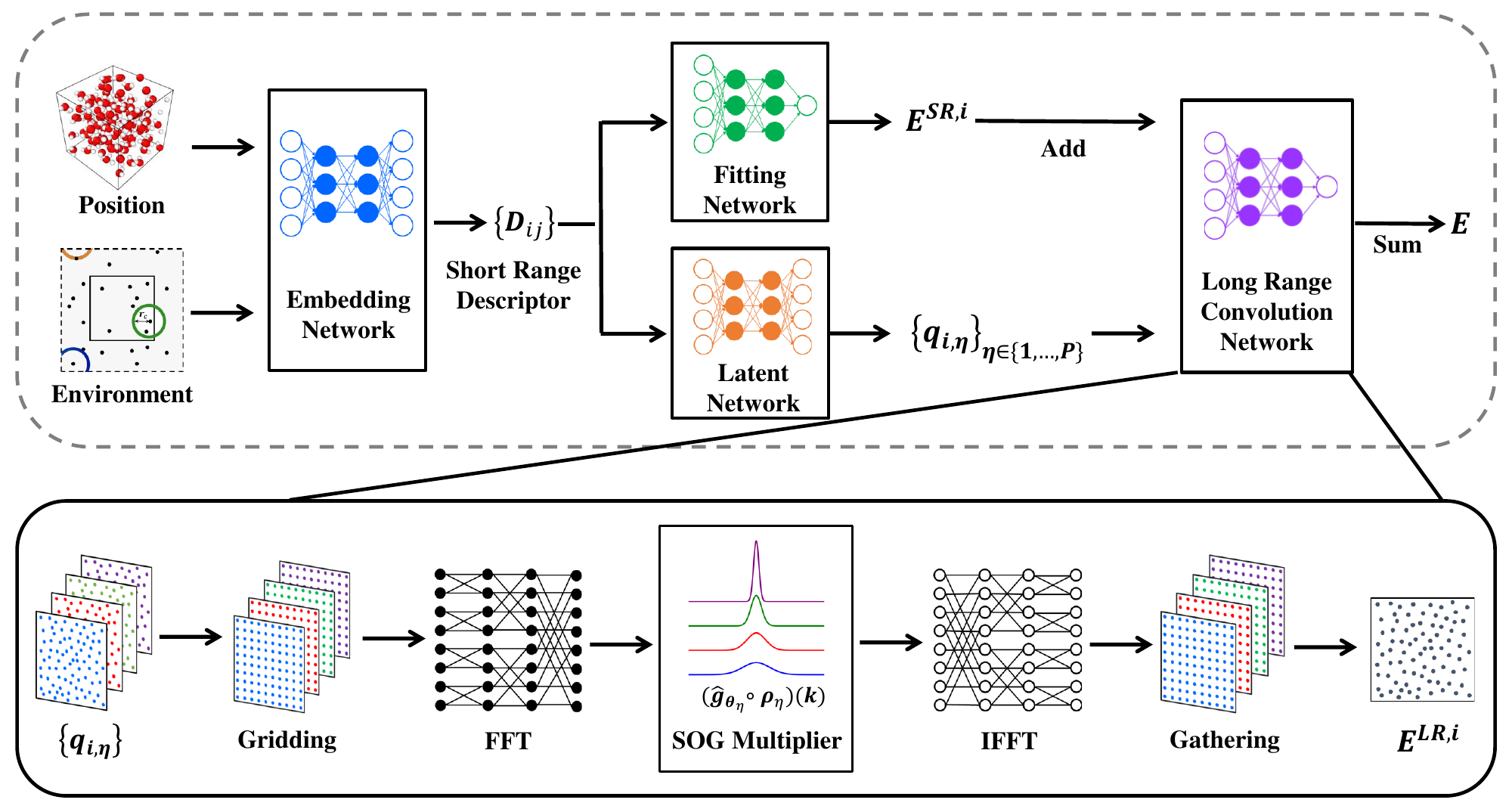}
    \caption{Schematic illustration of the SOG-Net model. The upper section depicts the entire network structure, whereas the lower section provides a detailed view of the LR convolution layer. Here, \(\{\bm{q}_{i,\eta}\}\), \(E^{SR,i}\) and \(E^{LR,i}\) represent the latent variables, SR and LR components of atomic energy, respectively.}
    \label{fig:3DSketch}
\end{figure}

Consider a system of $N$ atoms at $\bm{r}_{i}$, for $i=1, \cdots, N$, in a three-dimensional space with periodic boundary conditions. One assumes that the total potential energy can be written into SR and LR parts, 
$E=E^{SR}+E^{LR}.$
As is standard in most MLIPs~\cite{ko2023recent}, the SR energy is summed over the atomic contribution of each atom:
\begin{equation}
E^{SR}=\sum_{i=1}^{N}E^{SR,i}=\sum_{i=1}^{N}f_{\bm{\theta}_{SR}}(\bm{D}_{i}),
\end{equation}
where $\bm{D}_i$ is the SR descriptor, comprising invariant features of the $i$-th atom. The neural network $f_{\bm{\theta}_{SR}}(\cdot)$, parameterized by $\bm{\theta}_{SR}$, maps $\bm{D}_i$ to the atomic SR energy $E^{SR,i}$. The descriptor \(\bm{D}_i\) is a function of the local atomic environment $\bm{\mathcal{R}}_i \in \mathbb{R}^{\mathcal{N}_i \times 3}$, where $\mathcal{N}_i$ represents the number of neighbors of atom $i$ within a predefined cutoff $r_c$. To ensure physical consistency, $\bm{D}_i$ preserves symmetries such as translation, rotation and permutation. Further details on its practical construction are provided in Section \ref{subsec::SR} of the Supplementary Material (SM)~\cite{supplementary_information}.

Leveraging the rapid decay of LR interactions in Fourier space, we model the LR energy using multiple Fourier convolution layers:
\begin{equation}\label{eq::long-rangepotential}
E^{LR}= \frac{1}{2V}\sum_{i=1}^{N}\sum_{\eta=1}^{P}q_{i,\eta}\sum_{\bm{k}}  \widehat{g}_{\bm{\theta}_{\eta}}(\bm{k})\widehat{\rho}_{\eta}(\bm{k})e^{i\bm{k}\cdot\bm{r}_i},
\end{equation}
where $P\in \mathbb{Z}^+$ is the total number of layers, \begin{equation}
\bm{q}_{\eta}=(q_{1,\eta},\cdots,q_{N,\eta}),\quad \eta=1,\cdots,P
\end{equation}
denotes the $\eta$-th layer of latent variables for the LR energy, $\widehat{g}_{\bm{\theta}_{\eta}}$ is the SOG multiplier operator parametrized by $\bm{\theta}_{\eta}$ for the $\eta$-th layer, $\bm{k}=2\pi (m_x/L_x, m_y/L_y,m_z/L_z)$ denotes the Fourier modes with $m_x,m_y,m_z\in\mathbb{Z}$, and  
\begin{equation}
\widehat{\rho}_{\eta}(\bm{k}) = \sum_{j=1}^{N} q_{j,\eta} e^{-i \bm{k} \cdot \bm{r}_j}
\end{equation}
denotes the structure factor of the latent variables for the $\eta$-th layer. 
In the LR model, the latent variables $\bm{q}_{\eta}$ in each layer can be interpreted as the weights assigned to each atom for different types of LR interactions, such as partial charges in electrostatic interactions or atomic sizes in dispersion interactions. These latent variables are obtained by mapping the invariant features $\bm{D}_i$ to $\bm{q}_{\eta}$ using a latent neural network parameterized by $\bm{\theta}_{lnn}$, i.e., $\{q_{i,\eta}\}_{\eta=1}^{P}=f_{\bm{\theta}_{lnn}}(\bm{D}_i)$. A schematic diagram of the basic network structure is presented in the upper half of Figure~\ref{fig:3DSketch}.

In our LR model, two key aspects are the parameterization of the SOG multiplier operator $\widehat{g}_{\bm{\theta}_{\eta}}$ and the efficiency in computing Eq.~\eqref{eq::long-rangepotential}. These factors will be addressed sequentially as follows. We parameterize $\widehat{g}_{\bm{\theta}_{\eta}}$ as a sum of Gaussians,
\begin{equation}\label{eq::FourierMultiplier}
\widehat{g}_{\bm{\theta}_{\eta}}(\bm{k}) = \sum_{\ell=1}^{M} w_{\ell,\eta} e^{-k^2 / s_{\ell,\eta}^2},
\end{equation}
where $\bm{\theta}_{\eta} = \{w_{\ell,\eta}, s_{\ell,\eta}\}_{\ell=1}^{M}$ is the set of trainable parameters. The SOG multipliers $\{\widehat{g}_{\bm{\theta}_{\eta}}\}_{\eta=1}^{P}$ serve as a multilayer approximation of the LR tail of the potential in Fourier space. Gaussians exhibit excellent symmetry and smoothness and are known to minimize the Heisenberg uncertainty principle for $L^2$ functions~\cite{folland1997uncertainty}. These properties allow SOG multipliers to achieve an optimal trade-off between spatial and frequency localization, facilitating the effective representation of LR tails. The form of LR energy, as constructed in Eqs.~\eqref{eq::long-rangepotential}-\eqref{eq::FourierMultiplier}, inherently preserves translation, rotation, and permutation invariants. Additional constraints can  be imposed either on the coefficients \( \bm{\theta}_{\eta} \) or by projecting the learned potential onto a set of symmetric basis functions~\cite{huguenin2023physics,faller2024density}. 
In practice, the bandwidths are initialized as
$s_{\ell,\eta} = \exp[{b_{\text{min}} + \ell(b_{\text{max}} - b_{\text{min}}) / M}]$  with a uniform weight $w_{\ell,\eta}=1$.  The logarithmically spacing bandwidths effectively capture the multiscale nature of LR potentials and significantly reduce the number of Gaussians. Furthermore, after completing the training, $M$ is further reduced using model reduction techniques~\cite{benner2015survey} to improve the prediction efficiency. 
Moreover, we propose a fast algorithm embedded in the SOG-Net such that near-optimal complexity and high scalability can be achieved; See Appendix~\ref{app::fastalgorithm}.

To validate the SOG-Net model, we start by investigating a toy model of NaCl electrolytes. This system  consists of $1000$ particles interacting through Coulomb and Lennard-Jones forces. The training set contains at most $4000$ configurations, and the test set includes $200$ configurations, both sampled every $1000$ steps from a long MD trajectory using LAMMPS~\cite{thompson2022lammps} in the NVT ensemble at $T=300~K$. Further details on the training setup for the SR component are provided in Section~\ref{app:trainMD} of~\cite{supplementary_information}. For the LR component, we use a single layer of latent variables ($P=1$) and six Gaussians ($M=6$) in the LR convolution layer. 

\begin{figure}[t!] 
    \centering    
    \includegraphics[width=0.48\textwidth]{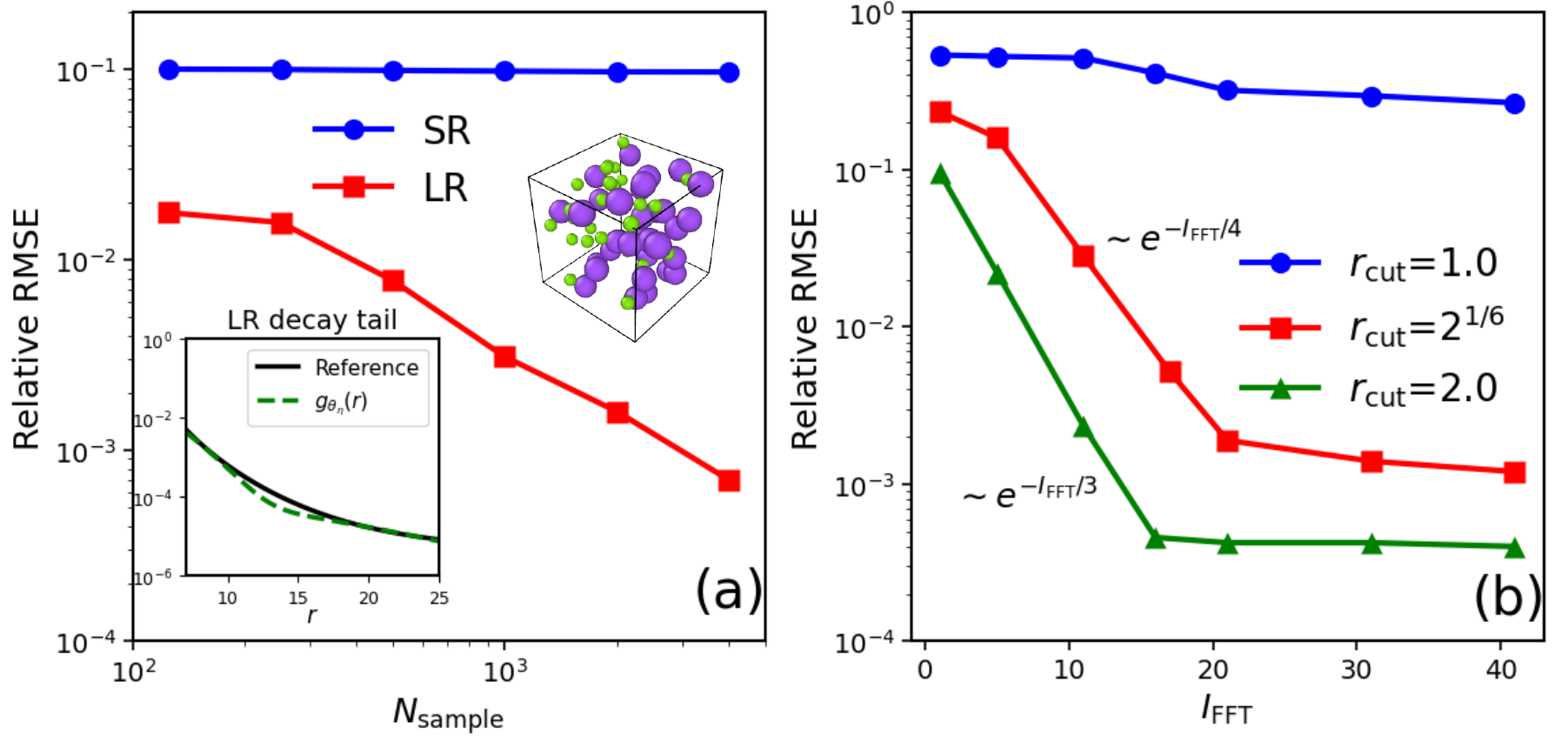}
    \caption{
(a) Test errors as a function of the size of training set. Data are shown for the short-range component (SR) and the full SOG-Net model (LR). The insets display a snapshot of the system and a plot of the learned SOG multiplier $g_{\bm{\theta}_{\eta}}$, respectively. The solid line represents the reference LR decaying tail in the force field.  (b) Test errors as a function of the number of Fourier grids \( I_{\text{FFT}} \). }
    \label{fig:3DFFTgrid}
\end{figure}

In Figure~\ref{fig:3DFFTgrid}(a), we plot the test error as a function of the training set size, keeping the number of Fourier grids fixed as $N_{\text{FFT}}=21^3$. One observes that using the SR part alone results in limitations in the representational capacity. However, combining the two components improves the accuracy by nearly 1–2 orders of magnitude. The inset on the bottom left of Figure~\ref{fig:3DFFTgrid}(a) demonstrates that our SOG multiplier effectively captures the LR tail, which is represented by a mix of electrostatic and Lennard-Jones interactions. In Figure~\ref{fig:3DFFTgrid}(b), we plot the test error of the complete network against the number of Fourier grids. When the cutoff radius of the SR network exceeds \(2^{1/6}\), the equilibrium distance for Lennard-Jones interactions, the SOG-Net achieves a relative error of \(10^{-3}\) using only $21$ grid points per dimension. Conversely, at a short cutoff radius ($r_c=1$), incorporating an LR layer fails to improve the fitting accuracy due to insufficient representation of repulsive interactions. These findings suggest that the choice of parameters for the SR and LR parts must be carefully balanced to optimize the accuracy. Furthermore, we test the SOG-Net on the same dataset with removing Coulomb interactions out and truncating Lennard-Jones interactions at $r_c = 2^{1/6}$, yielding a purely short-range system. The SOG-Net predicts negligible LR contributions, confirming that it does not overfit to noise (see Figure~\ref{fig:Pure} of \cite{supplementary_information} for the results). Additionally, we employed the benchmark for a molten bulk NaCl dataset from \cite{faller2024density} where results of many different models exist. The dataset is generated using PAW$\_$PBE (Na$\_$pv$/$Cl) pseudopotentials and features a LR decay of $1/r$. The SOG-Net achieves performance comparable to the corresponding Ewald-based CACE-LR result, both with and without message-passing layers, and outperforms other models (see Table~\ref{tab::TableComparison} in \cite{supplementary_information}).

\begin{figure}[!ht]
    \includegraphics[width=0.48\textwidth]{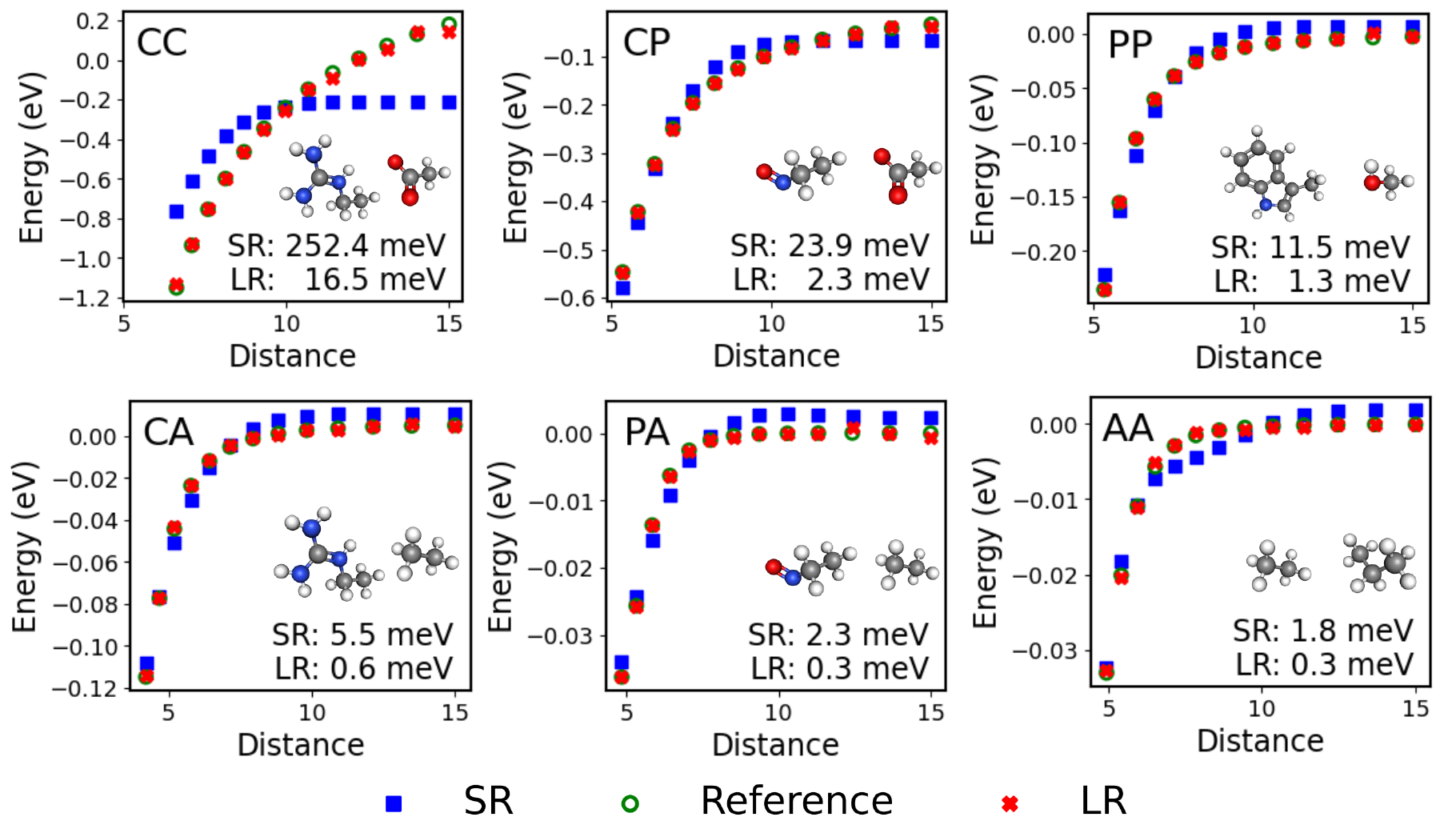}
    \caption{Comparison of the binding energy curves for the short-range component (SR) and the full SOG-Net model (LR) across six dimer classes: charged-charged (CC), charged-polar (CP), polar-polar (PP), charged-apolar (CA), polar-apolar (PA), and apolar-apolar (AA). The binding energy curve represents the potential energy difference between the dimer and its isolated monomers. Insets display  system snapshots with charge states labeled, along with the RMSE of energy for both models. }
    \label{fig:sixdimer}
\end{figure}

Next, we evaluate the SOG-Net by examining the binding curves of dimer pairs formed from charged (C), polar (P), and apolar (A) relaxed molecules at various separations (ranging from approximately $5\mathring{\text{A}}$ to $15\mathring{\text{A}}$) within a periodic cubic box of side length $L=30\mathring{\text{A}}$. These three molecular categories result in six distinct dimer classes (CC, CP, PP, CA, PA, and AA), characterized by ideal power-law decay constants $1/r^p$ with $p$ from 1 to 6. Snapshots of the dimer configurations are provided as insets in Figure~\ref{fig:sixdimer}. Our benchmark utilizes data from the BioFragment Database~\cite{burns2017biofragment}, computed using the HSE06 hybrid functional~\cite{heyd2003hybrid} with a high fraction of exact exchange. For each molecular pair, the dataset is randomly split into training and testing sets in a $9:1$ ratio. This benchmark is designed to evaluate the ability of the SOG-Net model to accurately capture the LR behavior of various interaction types.
In the LR component of the SOG-Net, we use a single layer of latent variables, $M=12$ Gaussians, and $N_{\text{FFT}}=31^3$ Fourier grids. The construction of the SR component and the training setup are detailed in Section~\ref{app:trainMD} of~\cite{supplementary_information}. Figure~\ref{fig:sixdimer} presents the binding energy curve. The pure SR models fail to accurately capture the training data, as evidenced by the noticeable flattening of the binding energy curves at larger separations between the two molecules. The improvement achieved by our LR models is particularly remarkable for the CA, PA, and AA cases, where the energy scales are much smaller (below $0.1$ meV). Similar performance for forces is shown in Figure~\ref{fig:SOG_Bar} in~\cite{supplementary_information}, where the force RMSEs (in meV/$\mathring{\text{A}}$) for the SR and full SOG-Net models across the whole dataset are  64.3 and 3.8 for CC,  35.4 and 2.1 for CP,  32.0 and 0.2 for PP, 26.8 and 0.5 for CA,  2.26 and 0.2 for PA, and  1.13 and 0.06 for AA, respectively. The results demonstrate that the model incorporating the LR component significantly outperforms the pure SR model in all cases, with the RMSE of the forces reduced by approximately an order of magnitude. These observations highlight the critical role of the LR component in accurately modeling atomic interactions.  

Below, we highlight the advantages of the SOG-Net compared to previous methods tested on the same dataset. In the generalized LODE method~\cite{huguenin2023physics}, the training accuracy depends heavily on guessing the potential exponent $p$ for the LR tail of each dimer class. In contrast, the SOG-Net eliminates the need for such manual tuning, as its SOG multiplier adaptively fits the LR tail during training. Additionally, the recently proposed latent Ewald method~\cite{cheng2024latent} also learns latent variables from invariant features but employs the Ewald summation for evaluating the LR energy. The latent Ewald approach is well-suited for interactions with a $1/r$ LR tail, offering benefits for the CC class; however, it yields only modest improvements for non-CC classes. The SOG-Net is able to accurately capture diverse LR tails across both CC and five non-CC cases. 
For example, in the CC, CP, and PP classes, the force RMSEs are $3.8$, $2.1$, and $0.2$ meV/$\mathring{\text{A}}$, respectively, compared to approximately $18$, $20$, and $5$ meV for the LODE method (with optimal $p$)~\cite{huguenin2023physics} and $5.9$, $16.5$, and $3.0$ meV for the latent Ewald approach~\cite{cheng2024latent,kim2024learning}. These results demonstrate the outstanding performance of the SOG-Net.


\begin{figure}[t!]
    \centering    \includegraphics[width=0.48\textwidth]{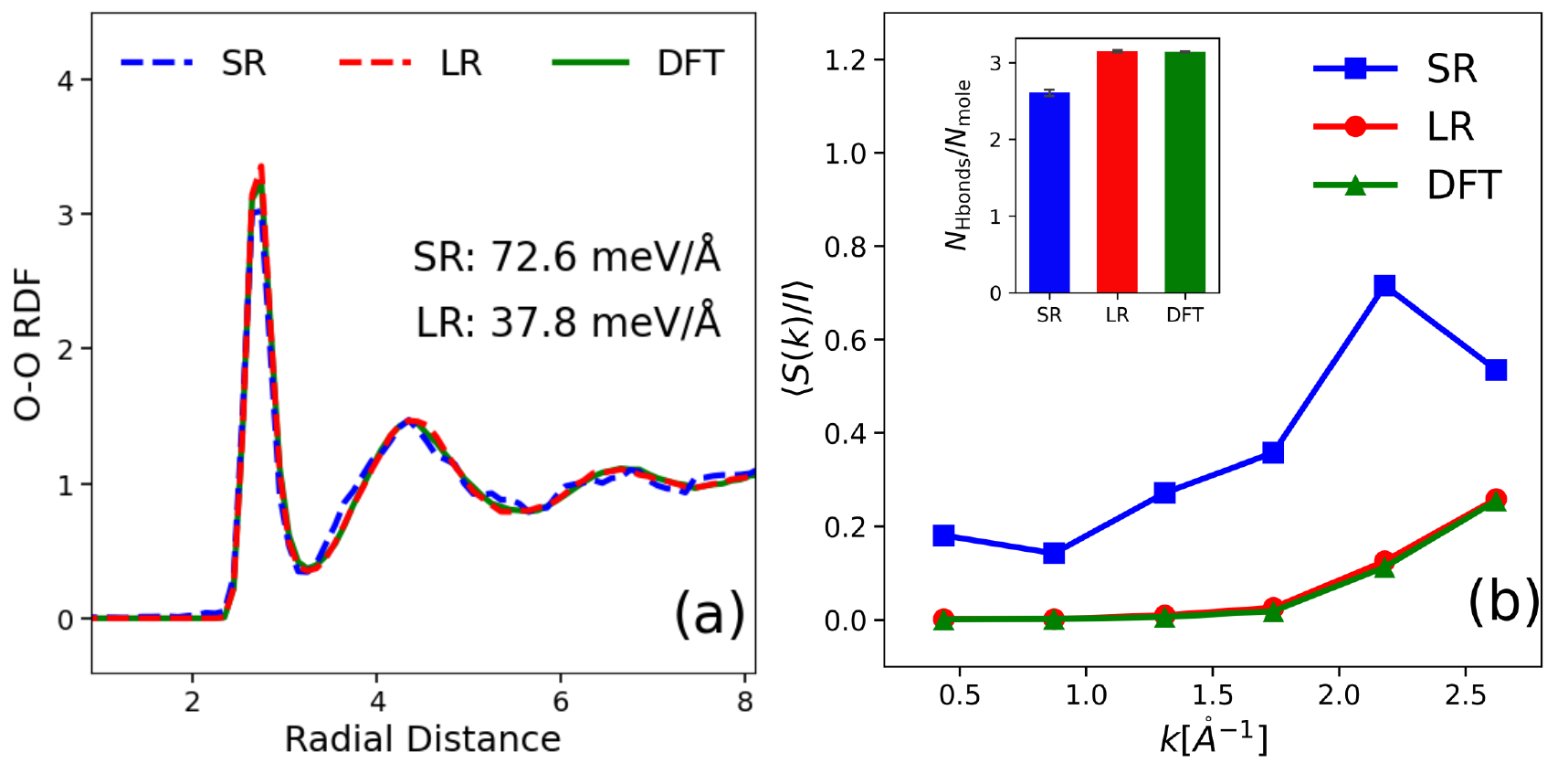}
    \caption{(a) Predicted radial distribution functions (RDFs) of oxygen-oxygen (O-O) atom pairs of water at $T=300 ~K$ and $1 g/mL$, using SR or LR models compared with the DFT results. (b) Predicted charge-charge structure correlations over charge density in reciprocal
    space as a function of $k$ values, using SR or LR models. The inset presents a plot of the average number of hydrogen bonds per water molecule.}
    \label{fig:water_rdf}
\end{figure}

Next, we applied the SOG-Net to a dataset of $1900$ liquid-water configurations, each containing $300$ atoms. The system has a density of $1~g/ml$ at $T = 300~K$. The dataset was generated via DFT calculations using the projector-augmented wave method~\cite{kresse1996efficient}. The electronic exchange-correlation energy was described using the generalized gradient approximation with the Perdew–Burke–Ernzerhof functional~\cite{payne1992iterative}. Details of the training setup for the SR component are provided in Section~\ref{app:trainMD} of~\cite{supplementary_information}. For the LR component, we use $P=16$ layers of latent variables, $M=12$ Gaussians for the SOG multiplier, and $31$ Fourier modes for each direction. The learning curves of the SOG-Net, shown in Figure~\ref{fig:SOG_Loss} in ~\cite{supplementary_information}, demonstrate that incorporating the LR component into the SOG-Net significantly reduces the error compared with the pure SR component. 

We performed MD simulations under the NVT ensemble. Figure~\ref{fig:water_rdf}(a) displays the oxygen-oxygen radial distribution function (RDF). RDFs for the oxygen-hydrogen and hydrogen-hydrogen pairs are provided in~\cite{supplementary_information} (Figure~\ref{fig:RDF_rdf}). All the computed RDFs are nearly identical and show excellent agreement with the DFT results. 
Figure~\ref{fig:water_rdf}(b) shows the longitudinal component of the charge structure factor along the $z$-axis, defined as
\begin{equation}
S(\bm{k})=\frac{1}{V}\sum_{ij=1}^{N}q_iq_j\langle e^{i\bm{k}\cdot \bm{r}_{ij}}\rangle,
\end{equation}
where $\langle\cdot\rangle$ denotes the ensemble average. The charge structure factor in the long-wavelength limit reflects the accuracy of reproducing LR correlations in water~\cite{hu2022symmetry}. The results show that our LR model well describes the charge structure factors, whereas pure SR models diverge sharply as $k\rightarrow 0$, consistent with observations in prior studies on SR and LR electrostatics~\cite{cox2020dielectric,schlaich2016water}. Additionally, we compute the average number of hydrogen bonds per water molecule using the pure SR model and the full SOG-Net model, as shown in the inset of Figure~\ref{fig:water_rdf}(b). The results confirm that incorporating the LR component significantly enhances the accuracy.
The time performance of the SOG-Net during MD simulations is given in Figure~\ref{fig:enter-label} of~\cite{supplementary_information}, demonstrating the linear scaling of the model. Although the theoretical complexity of the FFT and IFFT operators is $O(N\log N)$, their contribution to the overall computation time in the LR convolution layer is minimal. The CPU time of the full SOG-Net model is only slightly higher ($10\%-40\%$, depending on the system size) than that of the pure SR model, indicating that SOG-Net achieves significant improvements in fitting accuracy with only a modest increase in computational cost.

\begin{table}[ht]
\caption{RMSEs of energies (in meV/atom) and forces (in meV/$\mathring{\text{A}}$) for the SOG-Net in comparison with CC-ACE, 4G-HDNNP and CACE-LR.}
\label{tab::organicmolecule}
\centering
\resizebox{0.48\textwidth}{!}{
\begin{tabular}{|c|c|c|c|c|c|c|}
\hline
 & & CC-ACE & 4G-HDNNP & CACE-LR& SOG-Net \\
\hline
\multirow{3}{*}{C$_{10}$H$_2$/C$_{10}$H$_3^+$} 
& $r_{\text{cut}}$ & 6\,$\mathring{\text{A}}$  & 4.23\,$\mathring{\text{A}}$ & 4.23\,$\mathring{\text{A}}$& 4.23\,$\mathring{\text{A}}$ \\
& E &  0.75  & 1.194 & 0.73 & \textbf{0.67} \\
& F &  35.16  & 78.00 & 36.9 &\textbf{33.5}\\
\hline
\multirow{3}{*}{Na$_{8/9}$Cl$_{8}^+$}
& $r_{\text{cut}}$  & 6\,$\mathring{\text{A}}$  & 5.29\,$\mathring{\text{A}}$ & 5.29\,$\mathring{\text{A}}$ & 5.29\,$\mathring{\text{A}}$\\
& E &  0.71  & 0.481 & 0.21 & \textbf{0.124} \\
& F &  12.35  & 32.78 & 9.78 &\textbf{4.02}\\
\hline
\multirow{3}{*}{Au$_2$-MgO(001)}
& $r_{\text{cut}}$  & 6\,$\mathring{\text{A}}$  & 4.23\,$\mathring{\text{A}}$ & 5.5\,$\mathring{\text{A}}$ & 5.5\,$\mathring{\text{A}}$\\
& E  & 1.63 & 0.219 & 0.073 & \textbf{0.071} \\
& F  & 50.27  & 66.00 & 7.91 &\textbf{6.67}\\
\hline
\end{tabular}
}
\end{table}

Finally, we benchmark the SOG-Net on more complex datasets \cite{ko2021fourth}, including C$_{10}$H$_2$/C$_{10}$H$_3^+$, Na$_{8/9}$Cl$_{8}^{+}$, and Au$_2$ on MgO(001), where LR interactions and charge transfer effect are non-negligible. The SOG-Net training takes $P=1$, $M=12$, and $N_{\text{FFT}}=31^3$, with more detailed setup provided in Section IV of~\cite{supplementary_information}. Table~\ref{tab::organicmolecule} shows that the SOG-Net achieves significantly smaller errors among the compared models including charge-constrained ACE (CC-ACE)~\cite{rinaldi2025charge}, 4G-HDNNP~\cite{ko2021fourth} and CACE-LR~\cite{kim2024learning}. Additional tests on Au$_2$ on MgO(001) (see Appendix~\ref{app::AdditionalResults} for results) reveal that the SOG-Net accurately captures the energy differences between wetting and non-wetting states on Al-doped and undoped MgO surfaces, and resolves the distinct equilibrium bond lengths for different substrates, while SR models like 2G-HDNNP and CACE-SR fail to reproduce these LR charge transfer effects. 

On the physical interpretability of the SOG-Net model, the combination of latent variables and SOG multipliers can be viewed as a coarse-grained representation of electron density polarization. In the case of $P=1$ for a system with $1/r$ tail, the latent variables and SOG multiplier can be interpreted as atomic partial charges (similar to observations in \cite{kim2024learning} for Ewald-based method) and an SOG approximation of LR tail in Fourier space, respectively. The results on the SPICE dataset of polar dipeptide~\cite{eastman2023spice} demonstrate that SOG-Net can recover physically meaningful partial charges and multipoles of this case, achieving smaller validation errors than other models (see Figure \ref{fig:LRtransfer} and Table~\ref{tab:my_labelAccuracy} of \cite{supplementary_information}). For $P>1$, 
the network model gains capability for  better energy and force prediction in spite that one cannot perform explicit multipole extraction. 
Moreover, one may introduce the initial multipliers with an SOG approximation of a known physical kernel, or introduce penalty terms to enforce charge conservation, such that the stability and the extrapolation accuracy can be improved. We remark that the SOG-Net of the current version cannot explicitly handle three-body or higher-order LR interactions, which is expected to be addressed by integrating the network with the LODE framework~\cite{huguenin2023physics}. These directions will be explored in future work.



In conclusion, we have developed an efficient and versatile framework, SOG-Net, for modeling LR interactions in atomistic systems. Unlike existing LR methods, the SOG-Net requires no user-defined electrostatics or dispersion baseline corrections and does not rely on the classical Ewald summation, which assumes a predefined decay rate of $1/r$. It exhibits spectral convergence in learning and testing errors with the number of Fourier modes. Moreover, the LR MLIP incurs only a modest computational overhead compared to the pure SR network. This cost can be further reduced for both SR and LR components by leveraging acceleration techniques, such as random batch-type schemes~\cite{liang2021random,liang2023random}. By addressing these challenges, the SOG-Net provides a robust foundation for large-scale simulations with first-principles accuracy that account for LR interactions.
Notably, the LR convolution layer in the SOG-Net can also be integrated into existing methods such as HDNNP~\cite{behler2007generalized}, ACE~\cite{drautz2019atomic}, MACE~\cite{batatia2022mace}, CACE~\cite{cheng2024cartesian}, GAP~\cite{bartok2010gaussian}, and MTPs~\cite{shapeev2016moment}. In these approaches, after latent variables, such as partial charges, Wannier centers, or electronegativity, are learned through existing frameworks~\cite{unke2019physnet,ko2021fourth,gao2022self,zhang2022deep,rappe1991charge}, the SOG-Net can be readily integrated to enable adaptive learning of LR decay tails. This integration provides a simple mathematical formulation and achieves near-optimal computational complexity. 
The SOG-Net code is available on GitHub at https://github.com/DuktigYajie/SOG-Net.

\bigskip
\textit{Acknowledgments}---This work was supported by the National Natural Science Foundation of China (Grants No. 12325113, 12401570, 12426304, 124B2023 and 12350710181) and the Science and Technology Commission of Shanghai Municipality (Grant No. 23JC1402300). The work of J.L. is partially supported by the China Postdoctoral Science Foundation (grant No. 2024M751948). The authors would like to thank the support from the SJTU Kunpeng \& Ascend Center of Excellence.

\textit{Data availability}---The data that support the findings of this paper are openly available~\cite{SOG-Ne}.

\hspace{0.3mm}


\centerline{\textbf{\large End Matter}}
\appendix

\section{Fast algorithm in the SOG-Net}~\label{app::fastalgorithm}

Electrostatics is the bottleneck of MD with classical force fields. Many fast methods have been proposed for modeling and accelerating long-range electrostatics in the literature, including particle-mesh Ewald methods~\cite{Darden1993JCP,Hockney1988Computer}, fast multipole methods~\cite{greengard1987fast,greengard1988}, random batch methods \cite{jin2021random},  real-space truncation approaches \cite{wolf1999exact,fennell2006ewald}, and reaction-field models~\cite{barker1973monte}. These techniques can in principle be introduced to speed up the machine-learned long-range force fields. The SOG is an efficient kernel approximation method that describes diverse long-range decay behaviors well. It also allows the design of fast algorithms to achieve close-to-linear computational complexity considering that the Fourier transform of a Gaussian remains a Gaussian. Figure \ref{fig:3DSketch} presents the schematic of the fast method embedded in the long-range network. It  begins by mapping each layer of latent variables, $q_{i,\eta}$, onto a uniform grid through convolution with a shape function $W$, referred to as the gridding operator $\mathcal{G}_{\eta}$. The FFT operator $\mathcal{F}$ is then applied to compute $\widehat{\rho}_{\eta}(\bm{k})$. For each layer, the diagonal scaling operator $\mathcal{S}_{\eta}$  performs pointwise multiplication $(\widehat{g}_{\bm{\theta}_{\eta}}\circ \widehat{\rho}_{\eta})(\bm{k})$ in the Fourier space. Subsequently, the inverse FFT operator $\mathcal{F}^{-1}$ recovers the atomic LR energy at each grid point. Finally, the gathering operator $\mathcal{G}^{-1}_{\eta}$ maps the atomic energy back to the atomic positions $\{\bm{r}_i\}_{i=1}^{N}$, summing contributions across all layers. It is worth noting that the gridding, scaling, and gathering operators can be applied species-wise in practical implementations. The entire process of the LR convolution layer is summarized as
\begin{equation}\label{eq::operator}
E^{LR,i}=\mathcal{G}^{-1}_{\eta}\mathcal{F}^{-1}\mathcal{S}_{\eta}\mathcal{F}\mathcal{G}_{\eta}{\Big[}\{\bm{q}_{i,\eta}\}_{\eta=1}^{P}{\Big]},
\end{equation}
with a schematic illustration provided in the lower half of Figure~\ref{fig:3DSketch}. The detailed method is provided in Section~\ref{sec::FFTLR} of~\cite{supplementary_information}, where we leverage the FINUFFT package~\cite{Barnett2019SISC} to accelerate these operators via NUFFTs. As a result, each operator in Eq.~\eqref{eq::operator} can be executed in a complexity of either $O(N+N_{\text{FFT}}\log N_{\text{FFT}})$, where $N_{\text{FFT}}$ represents the average number of grid points per layer. This highlights a key advantage of the SOG-Net: after the learning process is complete, the model inherently supports acceleration through a fast algorithm, thereby enhancing the simulation process.

\hspace{3mm}

\section{Long-range charge transfer}\label{app::AdditionalResults}

\begin{figure}[!htbp]
 \centering  \includegraphics[width=0.96\linewidth]{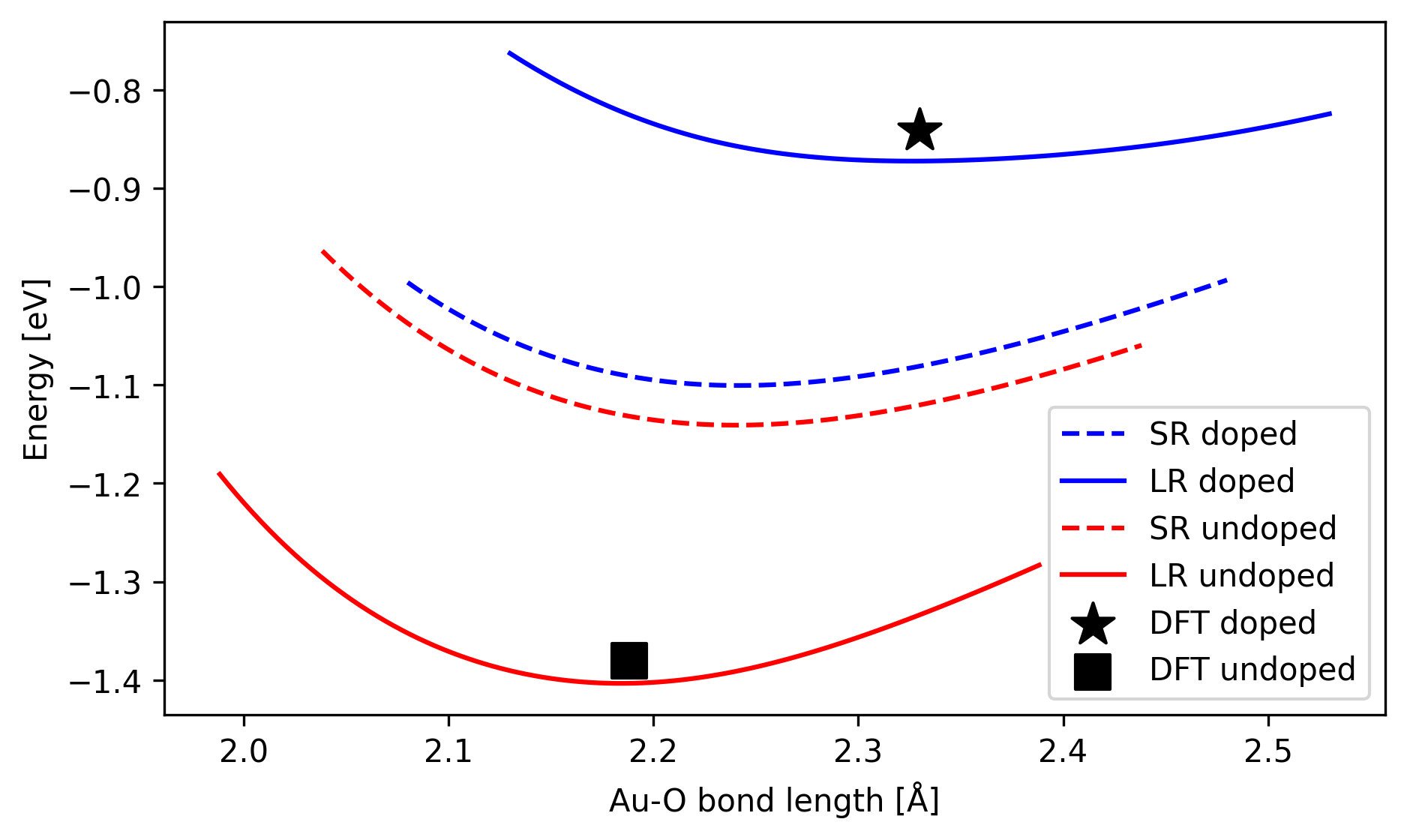}
\caption{Potential energy of the Au$_2$ clusters adsorbed on MgO(001) in the non-wetting geometry, for both Al-doped and undoped surfaces. Data are shown for results obtained using the pure short-range (SR) component and the full SOG-Net (LR). Equilibrium DFT bond lengths and corresponding minimum energies are marked by black symbols (star for doped, cube for undoped). The Au-O bond length is the minimum distance between Au and O atoms.}
\label{fig:enter-pot}
\end{figure}

This Appendix presents additional results on the ability of the SOG-Net in capturing long-range charge transfer in the Au$_2$-MgO(001) system. This system features a gold dimer adsorbed in two configurations: an upright, non-wetting state bonded to a surface O atom and a parallel, wetting state above two Mg atoms. Three Al dopants are placed in the fifth subsurface layer, significantly altering the PES between the two states. As shown in Table~\ref{tab:my_label}, the SOG-Net achieves lower errors than other methods. We further perform geometry optimizations of the gold atoms with the substrate fixed, following~\cite{ko2021fourth,kim2024learning}. Figure~\ref{fig:enter-pot} shows that the SOG-Net accurately resolves distinct equilibrium bond lengths for doped and undoped surfaces, highlighting its extrapolation ability in complex LR systems.

\onecolumngrid
\hspace{0.4mm}
\begin{table}[!htbp]
\centering
\resizebox{0.8\textwidth}{!}{
\begin{tabular}{|c|c|c|c|c|c|c|}
\hline & \text { DFT } & \text { 2G-HDNNP\cite{behler2007generalized} } & \text { CACE-SR\cite{cheng2024cartesian} } & \text { 4G-HDNNP\cite{ko2021fourth} } & \text { CACE-LR\cite{cheng2024latent} } & SOG-Net \\
\hline \text { Doped } & -66.9 & 375 & 431 & -41 & -70.6 & -67.0 \\
\text { Undoped } & 934.8 & 375 & 431 & 975 & 931.3 &931.9\\
\hline
\end{tabular}
}
\caption{Energy difference ($E_{\text{wetting}} - E_{\text{nonwetting}}$, in meV) between the wetting and non-wetting configurations for doped and undoped MgO(001) substrates. Data are shown for two SR models (2G-HDNNP\cite{behler2007generalized} and CACE-SR\cite{cheng2024cartesian}) and three LR models (4G-HDNNP\cite{ko2021fourth}, CACE-LR\cite{cheng2024latent} and SOG-Net).}
\label{tab:my_label}
\end{table}
\onecolumngrid
\clearpage
\onecolumngrid

\begin{center}
\textbf{\large Supplemental Material}
\end{center}
\renewcommand{\appendixname}{}  
\setcounter{section}{0}
\setcounter{figure}{0}
\setcounter{table}{0}
\setcounter{equation}{0}
\renewcommand{\thesection}{S\arabic{section}}
\renewcommand{\thefigure}{S\arabic{figure}}
\renewcommand{\thetable}{S\arabic{table}}
\renewcommand{\theequation}{S\arabic{equation}}

\section{Methods for the long-range network}\label{sec:method}
In this section, we discuss the motivation and technical details of the long-range (LR) convolution layer used in our sum-of-Gaussians neural network (SOG-Net), including the derivation of the LR model, the scheme for the parameter selection of Gaussian multipliers, and the implementation details of the trained model. 

\subsection{Limitations on using the Ewald summation in MLIPs}\label{subsec:Limitations}
Our primary motivation for developing the SOG-Net model is to enhance the widely used Ewald summation framework in machine learning interatomic potentials (MLIPs). In this subsection, we first discuss why the Ewald summation framework is unsuitable for LR tails that do not follow the $1/r^p$
form. We then introduce our improved model to address these limitations.

In classical models, charged particles are often modeled as point charges, with pairwise interactions following the empirical Coulomb form \(1/r\). Due to the long-range nature, direct computation has an unacceptably high complexity of \(O(N^2)\). To accelerate calculations, the Coulomb potential is often decomposed by Ewald splitting into the form
\begin{equation}
\frac{1}{r}=\frac{\text{erfc}(\alpha r)}{r}+\frac{\text{erf}(\alpha r)}{r},
\end{equation}
where $\text{erf}(\cdot)$ and $\text{erfc}(\cdot)$ represent the error function and complementary error function, respectively, and \(\alpha\) is known as the Ewald splitting parameter. The $\text{erfc}(\alpha r)/r$ term decays rapidly in real space and can thus be truncated for efficient computation. The $\text{erf}(\alpha r)/r$ term is singular in real space but very smooth, with
\begin{equation}\label{eq::1.3}
\frac{\text{erf}(\alpha r)}{r}\rightarrow \frac{1}{r},\quad \text{as}\quad r\rightarrow \infty \quad \text{and}\quad \alpha\in(0,\infty),
\end{equation}
making it suitable for truncation in Fourier space. Suppose that $N$ particles, $\{\bm{r}_i\}^{N}_{i=1}$, are located within a cubic box $\Omega=[0,L]^3$ with weights $\{q_i\}_{i=1}^{N}$. The periodic boundary conditions are imposed. The contributions on potential from the $\text{erf}(\alpha r)/r$ term is given by
\begin{equation}
\Phi(\bm{r}_i)=\sum_{\bm{n}\in\mathbb{Z}^3}\sum_{j=1}^{N}{}^{\prime}\frac{q_j \erf{(\alpha|\bm{r}_{ij}+L\bm{n}|)}}{|\bm{r}_{ij}+L\bm{n}|},
\end{equation}
where $\bm{r}_{ij}=\bm{r}_i-\bm{r}_j$ and the prime indicates that the case $i=j$ and $\bm{n}=0$ is excluded from the summation. Its Fourier spectral expansion can be expressed as
\begin{equation}\label{eq::1.4}
\Phi(\bm{r}_i)=\frac{2\pi}{V}\sum_{\bm{k}\neq \bm{0}}\frac{e^{-|\bm{k}|^2/(4\alpha^2)}}{|\bm{k}|^2}|\rho(\bm{k})|^2,
\end{equation}
where $\bm{k}=2\pi\bm{m}/L$, $\bm{m}\in\mathbb{Z}^3$, $\rho(\bm{k}):=\sum\limits_{j=1}^{N}q_{j}e^{i\bm{k}\cdot\bm{r}_j}$ denotes the charge structure factor, and $V=L^3$ denotes the volume of $\Omega$. 

Unlike classical models, molecules in density functional theory (DFT) simulations are not ideal point particles. Discrepancies arise due to spurious interactions between periodic images, particularly for charged fragments, and noise in binding energies at large distances, especially for weak interactions. Consequently, the LR decay rate of charge-charge interactions can exceed the classical $1/r$ Coulomb potential~\cite{huguenin2023physics}. Other LR interactions, such as dispersion and dipole-dipole interactions, may also deviate from their idealized decay rates. These characteristics are likely to be inherited by MLIPs since they are trained on DFT data. 


In the field of MLIP, Ewald summation is a commonly used approach for incorporating LR interactions. Two primary strategies have been developed: the first involves learning localized Wannier centers and introducing virtual charge sites, while the second focuses on predicting partial charges for each site while maintaining a fixed number of sites~\cite{zhang2022deep,ko2021fourth,shaidu2024incorporating}. Once these ``latent variables'' are inferred by the neural network, the Fourier components of the Ewald summation formula, Eq.~\eqref{eq::1.4}, are employed to calculate the contribution of charges to the potential energy. Recent graph-based MLIP studies have also utilized the Ewald summation for modeling LR message passing~\cite{kosmala2023ewald}. However, the use of Ewald summation, as described in Eq.~\eqref{eq::1.4}, assumes a predefined LR decay rate of $1/r$, derived from the $\erf(\alpha r)/r$ term in Eq.~\eqref{eq::1.3}. This fixed assumption may lead to systematic errors in MLIPs when the actual LR decay rate deviates from $1/r$.

\subsection{Derivation of long-range model}\label{subsec:Derivation}
To address the limitations of existing MLIPs, we replace the Ewald summation with a multilevel LR convolution layer, where the kernel function of each level is learned through a fully connected neural network, enabling adaptive learning of the LR tail while maintaining computational efficiency. 

Consider the LR atomic energy in the form of
\begin{equation}\label{eq::Phi1}
E^{LR,i}=\sum_{\eta=1}^{P}q_{i,\eta}\sum_{\bm{n}\in\mathbb{Z}^d}\sum_{j=1}^{N}{}^{\prime}q_{j,\eta} g_{\bm{\theta}_{\eta}}(\bm{r}_{ij}+L\bm{n}),
\end{equation}
where $P$ is the total number of layers, $\{q_{i,\eta}\big{|}~i=1,\cdots,N\}$ are the latent variables of $\eta$th layer, and $g_{\bm{\theta}_{\eta}}(\cdot)$ is a kernel function parameterized by $\bm{\theta}_{\eta}$ and trained via a neural network. The detailed representation of $g_{\bm{\theta}_{\eta}}(\cdot)$ is provided in subsection~\ref{subsec::Fouriermulti}. The LR energy is defined as $E^{LR}=\frac{1}{2}\sum_{i=1}^{N}E^{LR,i}$, where the factor $1/2$ account
for double-counting of interaction energy. In this context, $g_{\bm{\theta}_{\eta}}$ and $q_{i,\eta}$ at different layers can be interpreted as representations of various LR interactions and their corresponding coefficients, respectively. For $P=1$, $q_{i,1}$ is analogous to environment-dependent partial charges on individual atoms. For $P>1$, the model can incorporate more general LR interactions, thereby enhancing the representation and expressiveness of LR effects in the MLIP.

Let us define the Fourier transform pair as
\begin{equation}
\widehat{f}(\bm{k})=\int_{\Omega}f(\bm{r})e^{-i\bm{k}\cdot\bm{r}}d\bm{r},\quad\text{and}\quad f(\bm{r})=\frac{1}{V}\sum_{\bm{k}}\widehat{f}(\bm{k})e^{i\bm{k}\cdot\bm{r}}.
\end{equation}
For a given function $f$ defined on $\mathbb{R}^3$, the Poisson summation formula states that
\begin{equation}    \sum_{\bm{n}\in\mathbb{Z}^d} f(\bm{r} + L\bm{n}) = \frac{1}{V} \sum_{\bm{k}} \widehat{f}(\bm{k}) e^{i \bm{k} \cdot \bm{r}}.
\end{equation} 
By substituting the Poission summation formula into Eq.~\eqref{eq::Phi1}, we derive the Fourier spectral expansion of $E^{LR,i}$:
\begin{equation}\label{eq::phiexpan}
E^{LR,i}=\frac{1}{V}\sum_{\eta=1}^{P}q_{i,\eta}\sum_{j=1}^{N}q_{j,\eta}\sum_{\bm{k}}\widehat{g}_{\bm{\theta}_{\eta}}(\bm{k})e^{i\bm{k}\cdot(\bm{r}_i-\bm{r}_j)}-E^{\text{self},i},
\end{equation}
where
\begin{equation}\label{eq::self}
E^{\text{self},i}:=\frac{1}{V}\sum_{\eta=1}^{P}|q_{i,\eta}|^2\sum_{\bm{k}}\widehat{g}_{\bm{\theta}_{\eta}}(\bm{k})
\end{equation}
denotes the unwanted self-interaction energy. 

Given the slowly varying nature of LR interactions, we model \( g_{\bm{\theta}_{\eta}}(\bm{r}) \) as a radially symmetric function, i.e., \( g_{\bm{\theta}_{\eta}}(\bm{r}) = g_{\bm{\theta}_{\eta}}(r) \) with \( r = |\bm{r}| \). Consequently, its Fourier transform, \( \widehat{g}_{\bm{\theta}_{\eta}}(\bm{k}) = \widehat{g}_{\bm{\theta}_{\eta}}(k) \), is also radially symmetric. We refer to \( \widehat{g}_{\bm{\theta}_{\eta}}(k) \) as the ``SOG multiplier'', with the parameter \( \bm{\theta}_{\eta} \) being trainable. Our goal is to develop an efficient and scalable descriptor for the LR component of MLIPs through the following two steps: (1) The short-range (SR) descriptor is fed into a fully connected neural network to generate the latent variables \( \{q_{i,\eta}\}_{i=1}^{N} \) for each layer \( \eta = 1, \cdots, P \). (2) These latent variables are substituted into Eq.~\eqref{eq::phiexpan} to compute \( \Phi(\bm{r}_i) \), performed using Fourier diagonal scaling with the multiplier $\widehat{g}_{\bm{\theta}_{\eta}}(k)$. While directly evaluating \( \Phi(\bm{r}_i) \) for all \( i = 1, \ldots, N \)  requires \( O(NN_{\text{tot}}) \) operations with \( N_{\text{tot}} \) the total number of Fourier grids, the LR convolution layer (detailed in Section~\ref{sec::FFTLR}) reduces this computation to near-linear complexity in both \( N \) and \( N_{\text{tot}} \).



\subsection{Parameterization of the SOG multiplier}
\label{subsec::Fouriermulti}

In this work, we parameterize the SOG multiplier $\widehat{g}_{\bm{\theta}_{\eta}}$ as
\begin{equation}\label{eq::multiplier}
\widehat{g}_{\bm{\theta}_{\eta}}(\bm{k})=\sum_{\ell=1}^{M}w_{\ell,\eta}e^{-k^2/s_{\ell,\eta}^2},
\end{equation}
where $\bm{\theta}_{\eta}=\{w_{\ell,\eta},s_{\ell,\eta}\}_{\ell=1}^{M}$ is the set of trainable parameters of the $\eta$th layer. We initialize the weights $\{w_{\ell,\eta}\}$ as matrices of all ones. The bandwidths are initialized as
\begin{equation}
s_{\ell,\eta}=e^{b_{\text{min}}^{\eta}+\ell (b_{\text{max}}^{\eta}-b_{\text{min}}^{\eta})/M},\quad\, \ell=1,\cdots,M,
\end{equation}
where $b_{\text{min}}^{\eta},b_{\text{max}}^{\eta}\in\mathbb{R}$, $b_{\text{min}}^{\eta}<b_{\text{max}}^{\eta}$ so that $s_{\ell,\eta}$ are \emph{logarithmically} equally spaced points lying on $[e^{b_{\text{min}}^{\eta}},e^{b_{\text{max}}^{\eta}}]$ for each $\eta$. Here, we use logarithmic spacing nodes instead of uniform spacing to capture the multiscale characteristics of LR potentials and reduce the number of Gaussians needed for representation. The periodicity leads to a simple rule to set $b_{\text{min}}^{\eta}$ apriori for a given error tolerance $\varepsilon>0$: 
\begin{equation}
b_{\text{min}}^{\eta}\geq  \log\left(2\pi\right)-\log L-\log\sqrt{\log\varepsilon^{-1}},
\end{equation}
where this formula ensures that the Gaussian with the smallest bandwidth in the SOG multiplier has its support almost exclusively at the zero mode in Fourier space. Assume that a cutoff radius $r_c$ is specified for the local environment while constructing the SR descriptor, we also have a simple formula to set $b_{\text{max}}^{\eta}$:
\begin{equation}
b_{\text{max}}^{\eta}\leq  \log(2r_c^{-1})+\log\log\varepsilon^{-1}
\end{equation}
so that $g_{\bm{\theta}_{\ell}}(\bm{r})\sim \varepsilon$ when $r=r_c$. This ensure that the Gaussian with the largest bandwidth in $g_{\bm{\theta}_{\ell}}$ has its support larger than $r_c$ in real space.

The advantages of using an SOG to represent the SOG multiplier are threefold. First, as a type of radial basis function, Gaussians are well-suited for approximating unknown symmetric and smooth LR potentials. Second, Gaussians are renowned for minimizing the Heisenberg uncertainty principle for $L^2$ functions~\cite{folland1997uncertainty}, achieving an optimal trade-off between spatial and frequency localization. Given that LR potentials frequently exhibit multi-scale characteristics,  meaning they have significant support in both real and Fourier spaces, the properties of Gaussian functions make them particularly well-suited for fitting such functions. 


The multiplier in Eq.~\eqref{eq::multiplier} is highly versatile and enables the imposition of specific constraints on the learned potential. 
In our construction, both the multiplier \( \widehat{g}_{\bm{\theta}_{\eta}} \) and its inverse Fourier transform \( g_{\bm{\theta}_{\eta}} \) are radially symmetric, inherently ensuring translation, inversion, and rotational invariance. Additional constraints can be imposed on the coefficients \( \bm{\theta}_{\eta} \) by either solving an equation for some certain coefficients after training the others or projecting the learned LR atomic energy $E^{LR,i}$ onto a set of symmetric basis functions defined within the local environment of each atom similar to some previous works~\cite{huguenin2023physics,faller2024density}. 


In comparison to the Gaussian Approximation Potential (GAP)~\cite{bartok2010gaussian}, introduced in early machine learning models, our work presents two key distinctions. First, GAP is primarily designed to fit SR interactions, while our SOG-Net is employed to learn LR interactions in Fourier space. Second, GAP determines the coefficients through Gaussian regression, whereas our method learns \( \bm{\theta}_{\eta} \) via a neural network.

\subsection{Model Reduction}\label{subsec::Modelreduction}

In practice, a key problem is how to determine the number of Gaussians $M$ in the SOG multiplier. At the start of training, we choose a large $M$ to ensure sufficient resolution. Following each training round, model reduction (MR) techniques~\cite{benner2015survey} are used to find the optimal number of Gaussians. The MR process starts by converting the multiplier of each layer into a sum of exponentials via the change of variable $K=k^2$ and applying the Laplace transform $\mathcal{L}[\cdot]$, leading to a sum-of-poles representation:
\begin{equation}\label{eq::LaplaceMR}
\mathscr{L}[\widehat{g}_{\bm{\theta}_{\eta}}(K)]=\mathscr{L}\left[\sum_{\ell=1}^{M}w_{\ell,\eta}e^{-K/s_{\ell,\eta}^2}\right]=\sum_{\ell=1}^{M}\frac{w_{\ell,\eta}}{z+1/s_{\ell,\eta}^2}=\bm{c}(z\bm{I}-A)^{-1}\bm{b},
\end{equation}
where $\bm{A}$ is an $(M+1)\times(M+1)$ diagonal matrix, $\bm{b}$ and $\bm{c}$ are column and row vectors of dimension $M+1$, respectively. Next, one regards the rightmost term in Eq.~\eqref{eq::LaplaceMR} as a transfer function of a linear dynamical system, enabling the use of the balanced truncation method to reduce the number of terms to $\widetilde{M}<M+1$. The initial number of Gaussians used before reduction is reported in the main text. For DFT datasets, we typically start with $M=12$ (and $M=6$ for the toy NaCl model). After reduction, $M$ is usually halved without loss of accuracy. The Python implementation of the reduction procedure is included in the released SOG-Net package~(https://github.com/DuktigYajie/SOG-Net). 

\section{Methods for the short-range network}\label{subsec::SR}
In this section, we discuss the construction of SR component of our SOG-Net as well as some implementation details which is useful for improve the efficiency. 

The first step in our model is to generate a robust SR descriptor that preserves symmetry. This descriptor is then fed into two  networks: the first, referred to as the ``fitting network'', generates the SR potential; the second, called the ``latent network'', learns the latent variables $\{q_{i,\eta}\}$ from the SR descriptor. Notably, SOG-Net can be seamlessly integrated with any existing methods for constructing SR descriptors, such as tensor operations in Deep Potential~\cite{zhang2018deep} and moment tensor networks~\cite{shapeev2016moment}, or basis projections used in MACE~\cite{batatia2022mace}, CACE~\cite{cheng2024cartesian}, and LODE~\cite{Grisafi2019JCP}. Unless otherwise noted, the SR descriptors used in the examples presented in this paper are based on tensor operations similar to those in Deep Potential, due to their simplicity of implementation. Exploring alternative SR descriptors and identifying the most suitable approach will be a focus of future work.   

Next, we briefly review the construction of the SR model. The SR component of energy, $E^{SR}$, is computed as the sum
of atomic contributions from all atoms:
\begin{equation}\label{eq::II1}
E^{SR}=\sum_{i=1}^{N}E^{SR,i}=\sum_{i=1}^{N}f_{\bm{\theta}_{fit}}(\bm{D}_i),
\end{equation}
where $f_{\bm{\theta}_{fit}}$ is a residual neural network (ResNet), referred to as the ``fitting network'', and $\bm{D}_i$ is a column vector that represents the local environment of atom $i$. The vector $\bm{D}_i$ is core-shelly structured as  $\bm{D}_i=\{\bm{D}_{i,c},\bm{D}_{i,s}\}$, where $\bm{D}_{i,c}$ and $\bm{D}_{i,s}$ contain information about neighboring atoms within specified cutoff radii $r_c$ and $r_s$, respectively, with $r_c<=r_s$. The details for constructing $\bm{D}_{i,c}$ and $\bm{D}_{i,s}$ are provided below. Please note that we may use only $\bm{D}_{i,c}$ or $\bm{D}_{i,s}$ in numerical tests, depending on the actual situation, as set up in Section~\ref{app:trainMD}. The SR component of force, $\bm{F}^{\text{SR},i}$, can be derived by applying the automatic differentiation on $E^{SR}$.

The core component of the SR descriptor, $\bm{D}_{i,c}$, is derived from the environment matrix $\bm{\mathcal{R}}_{i,c}$ and embedding matrix $\bm{G}_{i,c}$:
\begin{equation}\label{eq::SRCore}
\bm{D}_{i,c}=\frac{1}{N_c^2}(\bm{G}_{i,c})^T\bm{\mathcal{R}}_{i,c}(\bm{\mathcal{R}}_{i,c})^{T}\bm{G}_{i,c}^<,
\end{equation}
where $N_c$ denotes the maximum number of neighbors within the cutoff radius $r_c$, $\bm{G}_{i,c}$ and $\bm{\mathcal{R}}_{i,c}$ have dimensions  $N_{c} \times M_c$ and $N_c\times 4$, respectively. Each row in $\bm{\mathcal{R}}_{i,c}$ is a four-dimensional vector:
\begin{equation}
\bm{\mathcal{R}}_{i,c}=\left\{\text{soft}(r_{ij})\times \left[\frac{1}{r_{ij}},~\frac{x_{ij}}{r_{ij}^2},~\frac{y_{ij}}{r_{ij}^2},~\frac{z_{ij}}{r_{ij}^2}\right]\bigg{|}~j=1,\cdots,N_c~\right\},
\end{equation}
where $r_{ij}=|\bm{r}_{ij}|$ with $\bm{r}_{ij}:=\bm{r}_i-\bm{r}_j$ . Here, the soft function $\text{soft}(\cdot)$ is defined by \begin{equation}\label{eq::soft}
\text{soft}(r)=\begin{cases}
1,& r<j_cr_c,\\[1em]
-6\left(\dfrac{r-j_cr_c}{r_c-j_cr_c}\right)^5+15\left(\dfrac{r-j_cr_c}{r_c-j_cr_c}\right)^4-10\left(\dfrac{r-j_cr_c}{r_c-j_cr_c}\right)^3+1, & j_cr_c\leq r<r_c,\\[1em]
0, & r\geq r_c,
\end{cases}
\end{equation}
where $0<j_c<1$ is a scaling factor determining the smooth transition region. The function $\text{soft}(r)$ smoothly decays from $1$ to $0$ within the cutoff range $r_c$, ensuring that its second derivative is continuous. This property guarantees continuity in the computed forces, which significantly enhances the long-term stability of MD simulations. The embedding matrix $\bm{G}_{i,c}$ is derived from the output of a neural network 
\begin{equation}
\bm{G}_{i,c}=f_{\bm{\theta}_{emb,c}}\left(\left\{\frac{\text{soft}(R_{ij})}{|R_{ij}|}\bigg{|}~j=1,\cdots,N_c\right\}\right),
\end{equation}
where the fully connected embedding neural network $f_{\bm{\theta}_{emb,c}}$ maps $\text{soft}(R_{ij})/R_{ij}$ to an $M_c$ dimensional vector. A sub-matrix of $\bm{G}_{i,c}$ that includes its first $M_c^<$ columns, with $M_c^< < M_c$, is denoted by $\bm{G}_{i,c}^<$. The resulting core component of the SR descriptor from Eq.~\eqref{eq::SRCore}, $\bm{D}_{i,c}$, has dimension $M_c\times M_c^<$, encapsulates both angular and radial information, and preserves symmetry.

The shell component of the SR descriptor, $\bm{D}_{i,s}$, is defined as:
\begin{equation}
\bm{D}_{i,s}= \frac{1}{N_s} \sum_{j=1}^{N_s} (\bm{G}_{i,s})_j,
\end{equation}
where $N_s$ is the maximum number of neighboring atoms within the cutoff $r_s$, $\bm{G}_{i,s}\in \mathbb{R}^{N_s\times M_s}$ is an embedding matrix, and the summation aggregates the rows of $\sum_{j}$. For fewer than $N_s$ neighbors, the corresponding matrix terms are zero-padded. The embedding matrix $\bm{G}_{i,s}$ comprises $M_s$ nodes from the output layer of a neural network function $f_{\bm{\theta}_{emb,s}}$ applied to $\text{soft}(R_{ij})$:
\begin{equation}
(\bm{G}_{i,s})_j=f_{\bm{\theta}_{emb,s}}(\text{soft}(R_{ij})),
\end{equation}
which includes only radial neighbor information. The soft function, $\text{soft}(\cdot)$, is defined as in Eq.~\eqref{eq::soft}. In practice, we set $r_c<r_s$, since three-body interactions are more critical for close neighbors, while radial information alone suffices for distant neighbors.

After generating the core component $\bm{D}_{i,c}$ and shell component $\bm{D}_{i,s}$, $\bm{D}_{i,c}$ is flattened into a vector and concatenated with $\bm{D}_{i,s}$ to form the short-range descriptor $\bm{D}_i=\{\bm{D}_{i,c},\bm{D}_{i,s}\}$. The whole process involves constructing a stencil of bins to identify potential neighbors, binning atoms, and looping through the stencil to assemble the neighbor list $\mathcal{N}_i$ based on the core and shell cutoff radii $r_c$ and $r_s$. To improve efficiency, we employ the \(kd\)-tree algorithm, which uses hierarchical spatial decomposition to perform fast nearest-neighbor searches with linear complexity.

After generating the SR descriptor $\bm{D}_i$, it is fed into two separate networks. The first is the fitting network, which computes the SR atomic energy as described in Eq.~\eqref{eq::II1}; the second is a fully connected ``latent'' network, $f_{\bm{\theta}_{lnn}}(\cdot)$, which maps the input SR descriptor to output latent variables. These latent variables are then processed by the LR convolution layer to compute the LR atomic energy. The acceleration of this process is outlined in the next section.  

\section{FFT-accelerated fast long-range convolution}\label{sec::FFTLR}
In this section, we detail how the fast Fourier transform (FFT) is used to accelerate calculations for the LR convolution layer, as described in Eq.~\eqref{eq::phiexpan}, within our SOG-Net model. 
Although the FFT is a well-established approach, however, the non-uniform distribution of atoms in MD systems prevents the direct application of the FFT. Moreover, unlike the Ewald summation which interaction kernel is fully deterministic, our model employs the trainable kernel $\widehat{\bm{g}}_{\bm{\theta}_{\eta}}$, which evolves dynamically during the training process. Below, we propose a fast method to address these challenges.  


Let us start by introduce a localized, separable and periodic shape function $W(\bm{r})$ along with its Fourier transform $\widehat{W}(\bm{k})$ for spreading particles onto a uniform grid and collecting field values from the grid. By inserting the identity $1\equiv\widehat{W}(\bm{k})|\widehat{W}(\bm{k})|^{-2}\widehat{W}(\bm{k})$ into Eq.~\eqref{eq::phiexpan}, one has
\begin{equation}\label{eq::phiint}
E^{LR,i}=\frac{1}{V}\sum_{\eta=1}^{P}q_{i,\eta}\sum_{\bm{k}}\widehat{W}(\bm{k})e^{i\bm{k}\cdot\bm{r}}\frac{\widehat{g}_{\bm{\theta}_{\eta}}(\bm{k})}{|\widehat{W}(\bm{k})|^2}\sum_{j=1}^{N}q_{j,\eta}\widehat{W}(\bm{k})e^{-i\bm{k}\cdot\bm{r}_j}.
\end{equation}
We will use the last $\widehat{W}(\bm{k})$ for the gridding step, the first $\widehat{W}(\bm{k})$ for the gathering step, and $|\widehat{W}(\bm{k})|^{-2}$ for the diagonal scaling step, respectively. The details are provided as follows.

The calculation of Eq.~\eqref{eq::phiint} is performed level-by-level. For each level $\eta$, let us define 
\begin{equation}
\widehat{\Phi}_{\text{grid},\eta}(\bm{k}):=\sum_{j=1}^{N}q_{j,\eta}\widehat{W}(\bm{k})e^{-i\bm{k}\cdot\bm{r}_j},
\end{equation}
which is the Fourier transform of 
\begin{equation}\label{eq::grid}
\Phi_{\text{grid},\eta}(\bm{r})=\sum_{j=1}^{N}q_{j,\eta}W(\bm{r}-\bm{r}_j)
\end{equation}
by the convolution theorem. The evaluating of $\Phi_{\text{grid},\eta}$ on a uniform grid will be defined as the gridding operator $\mathcal{G}_{\eta}$. After this step is done, we obtain $\widehat{\Phi}_{\text{grid},\eta}$ from $\Phi_{\text{grid},\eta}$ by applying a 3D FFT operator $\mathcal{F}$. Next, we compute
\begin{equation}\label{eq::scal}
\widehat{\Phi}_{\text{scal},\eta}(\bm{k}):=\frac{\widehat{g}_{\bm{\theta}_{\eta}}(\bm{k})}{|\widehat{W}(\bm{k})|^2}\widehat{\Phi}_{\text{grid},\eta}(\bm{k})
\end{equation}
for each Fourier mode which is defined as the diagonal scaling operator $\mathcal{S}_{\eta}$ (here diagonal means without calculations for cross terms). Next, we obtain $\Phi_{\text{scal},\eta}$ on the real-space grid from $\widehat{\Phi}_{\text{scal},\eta}$ by applying an 3D inverse FFT (IFFT) operator $\mathcal{F}^{-1}$. Note that Eq.~\eqref{eq::scal} allow us to write Eq.~\eqref{eq::phiint} as
\begin{equation}\label{eq:Phigat}
E^{LR,i}=\sum_{\eta=1}^{P}q_{i,\eta}\int_{\Omega}\Phi_{\text{scal},\eta}(\bm{r}')W(\bm{r}_i-\bm{r}')d\bm{r}':=\sum_{\eta=1}^{P}q_{i,\eta}\Phi_{i,\eta}
\end{equation}
by applying again the convolution theorem. The gathering operator $\mathcal{G}^{-1}_{\eta}$ is performed by discretizing Eq.~\eqref{eq:Phigat} on a uniform grid using the trapezoidal rule and summing the contributions from grids for each atom. When the shape function $W$ is smooth and compactly supported, the algorithm achieves spectral convergence as the Fourier grid size increases due to the use of FFT. The entire process of the LR convolution layer is summarized as:
\begin{equation}
E^{LR,i}=\sum_{\eta=1}^{P}q_{i,\eta}\mathcal{G}^{-1}_{\eta}\mathcal{F}^{-1}\mathcal{S}_{\eta}\mathcal{F}\mathcal{G}_{\eta}\left[\{q_{j,\eta}\big{|}~j=1,\cdots,N;~\eta=1,\cdots,P\}\right].
\end{equation}
 It is worth noting that the gridding, scaling, and gathering operators can be applied species-wise for multi-species systems. Algorithm~\ref{al::mid} outlines the procedure for this fast algorithm, which significantly accelerates both the training and testing processes of our SOG-Net model.

In practice, these operators are efficiently implemented using  existing packages. The gridding and 3D FFT operators can be performed by Type-1 non-uniform FFTs (NUFFTs)~\cite{dutt1993fast,greengard2004accelerating}, while the 3D IFFT and gathering operators can be performed by Type-2 NUFFTs. The NUFFT is particularly effective for accelerating the exponential sums of the form $\sum_i w_i e^{i\bm{k}\cdot\bm{r}_i}$, where $\bm{k}$ and/or $\bm{r}_i$ exhibit non-uniform distributions. 
In this work, we utilize the FINUFFT package~\cite{Barnett2019SISC} to accelerate the NUFFT computations. The error tolerance is set to $10^{-4}$, and the so-called ``exponential of semicircle'' kernel is used as the shape function. The total computational complexity for each layer $\eta$ is $O(N+N_{\text{FFT},\eta}\log N_{\text{FFT},\eta})$, where $N_{\text{FFT},\eta}$ denotes the size of the Fourier grid for layer $\eta$.



\begin{algorithm}[H]
\caption{~Accelerating the long-range convolution layer in the SOG-Net model}\label{al::mid}
\begin{algorithmic}[1]
\Statex \textbf{Input:} Positions $\{\bm{r}_i\}$, the side length $L$, the shape function $W$, and the size of FFT grid $N_{\text{FFT},\eta}$, the latent variables $\{q_{i,\eta}\}$ and the SOG multipliers $\widehat{g}_{\bm{\theta}_{\eta}}$ at each layer $\eta=1,\cdots,P$. 
\State (Gridding) For each layer $\eta$, spreading latent variables from atomic locations to a uniform grid to obtain $\Phi_{\text{grid},\eta}$ by using Eq.~\eqref{eq::grid}.
\State (FFT) Obtain $\widehat{\Phi}_{\text{grid},\eta}$ on the corresponding Fourier grid from $\Phi_{\text{grid},\eta}$ via a 3D FFT.
\State (Diagonal Scaling) For each layer $\eta$, obtain $\widehat{\Phi}_{\text{scal},\eta}$ by scaling $\widehat{\Phi}_{\text{grid},\eta}$ with the SOG multiplier $\widehat{g}_{\bm{\theta}_{\eta}}$ and the $\widehat{W}^{-2}$ term as in Eq.~\eqref{eq::scal}.
\State (IFFT) Obtain $\Phi_{\text{scal},\eta}$ on real-space grids from $\widehat{\Phi}_{\text{scal},\eta}$ via a 3D IFFT.
\State (Gathering) For each layer $\eta$, obtain $\Phi_{i,\eta}$ by gathering contributions from grids as in Eq.~\eqref{eq:Phigat}.
\Statex \textbf{Output:} Long-range atomic energy $E^{LR,i}=\sum\limits_{\eta=1}^{P}q_{i,\eta}\Phi_{i,\eta}$.
\end{algorithmic}
\end{algorithm}	

The LR component of the force, $\bm{F}^{LR,i}$, can be derived by applying automatic differentiation to $E^{LR}$. However, applying automatic differentiation for FFT and IFFT operators in the current implementation of the FINUFFT package is of rather expensive. We instead compute the gradient of $E^{LR}$ with respect to the atomic position $\bm{r}_i$ directly using Eq.~\eqref{eq::Phi1}. The formula is expressed as:
\begin{equation}\label{eq::FLRi}
\bm{F}^{LR,i}=-\nabla_{\bm{r}_i}E^{LR}=\frac{1}{2V}\sum_{\eta=1}^{P}q_{i,\eta}\sum_{\bm{k}}\bm{k}\widehat{g}_{\bm{\theta}_{\eta}}(\bm{k})\text{Im}\left[e^{i\bm{k}\cdot\bm{r}_i}\sum_{j=1}^{N}q_{j,\eta}e^{-i\bm{k}\cdot\bm{r}_j}\right],
\end{equation}
where $\text{Im}[\cdot]$ denotes the imaginary part. In practice, the computation of Eq.~\eqref{eq::FLRi} can be also accelerated by utilizing three additional Type-2 NUFFTs during the gathering step.

\section{Training and simulation details}\label{app:trainMD}

In this section, we provide details on the training and simulation setups used in our tests. 

We start by describing the loss function and the error metrics used in our analysis. Given the importance of forces compared to energy in MD simulations, the loss function is defined as the mean squared error (MSE) of forces:
\begin{equation}\label{eq::loss}
\mathcal{E}_{\text{loss}}^{\text{force}}:=\frac{1}{N_{\text{batch}}}\sum_{n=1}^{N_{\text{batch}}}\sum_{i=1}^{N}\|\bm{F}^{\text{SOG-Net}}_{\text{tot},n}(\bm{r}_i)-\bm{F}_{\text{tot},n}^{\text{exact}}(\bm{r}_i)\|^2,
\end{equation}
where $n$ indexes the batches sampled from the training set in each epoch, \(N_{\text{batch}}\) is the total number of batches, and \(\bm{F}_{\text{tot},n}^{\text{SOG-Net}}\) and \(\bm{F}_{\text{tot},n}^{\text{exact}}\) represent the forces calculated by SOG-Net and the reference solution derived from DFT, respectively. For tests involving binding curves of dimers, the loss function also involves the MSE of energies:
\begin{equation}\label{eq::lossEner}
\mathcal{E}_{\text{loss}}^{\text{energy}}:=\frac{1}{N_{\text{batch}}}\sum_{n=1}^{N_{\text{batch}}}\|E^{\text{SOG-Net}}_{\text{tot},n}-E_{\text{tot},n}^{\text{exact}}\|^2.
\end{equation}
The loss function is given by $\mathcal{E}_{\text{loss}}=\mathcal{E}_{\text{loss}}^{\text{energy}}+\beta~\mathcal{E}_{\text{loss}}^{\text{force}}$ where $\beta$ is a non-negative factor. For dimer pairs, we set $\beta=1000$. More implementation details can be found in our code repository~(https://github.com/DuktigYajie/SOG-Net). 

During testing, the relative root-mean-square error (R-RMSE) of forces is used as the error metric for the NaCl electrolyte model. It is defined as:
\begin{equation}\label{eq::rel}
\mathcal{E}_{\text{R-RMSE}}:=\sqrt{\frac{\sum\limits_{m=1}^{N_{\text{test}}}\sum\limits_{i=1}^{N}\left\|\bm{F}_{\text{tot},m}^{\text{exact}}(\bm{r}_i)-\bm{F}_{\text{tot},m}^{\text{SOG-Net}}(\bm{r}_i)\right\|^2}{\sum\limits_{m=1}^{N_{\text{test}}}\sum\limits_{i=1}^{N}\left\|\bm{F}_{\text{tot},m}^{\text{exact}}(\bm{r}_i)\right\|^2}},
\end{equation}
where $m$ indexes the 
test configurations, and $N_{\text{test}}$ is the size of testing set. For tests involving binding energy curves, the RMSE of energies is also employed as an error metric. In other tests, we calculate the RMSE of forces across all test configurations as:
\begin{equation}
\mathcal{E}_{\text{RMSE}}^{\text{force}}:=\sqrt{\frac{1}{\text{NN}_{\text{test}}}\sum\limits_{m=1}^{N_{\text{test}}}\sum\limits_{i=1}^{N}\left\|\bm{F}_{\text{tot},m}^{\text{exact}}(\bm{r}_i)-\bm{F}_{\text{tot},m}^{\text{SOG-Net}}(\bm{r}_i)\right\|^2}.
\end{equation}

We next outline the network architecture and training approaches used in our tests. For the NaCl electrolyte model tests, we use only the shell component of the SR network described in Section~\ref{subsec::SR}, with cutoff radii set to $r_c=0$ and $r_s=2$ (adjusted as specified, if applicable). The sizes of the embedding network $f_{\bm{\theta}_{emb,s}}$ and the fitting network $f_{\bm{\theta}_{fit}}$ are configured as $(2,4,8,16,32)$ and $(32,32,32,32,32)$, respectively. The activation function after each layer is set as tanh to prevent gradient disappearance/explosion.
In this system, all ions are modeled as soft spheres of diameter $\sigma$ with $\sigma$ the distance unit, that interact through a force field that includes both the Coulomb potential $1/r$ and the dispersive Lennard-Jones (LJ) potential. Since the charge strengths are explicitly defined for these tests, with Na and Cl atoms assigned $+1$ and $-1$ charges, respectively, the latent network is not required to generate latent variables from the SR descriptor. Parameters for the LR component of SOG-Net model are detailed in the main text. The models are trained using the Adam optimizer with an initial learning rate of $0.001$, which decays by a factor of $0.97$ every $10$ epochs. The training process comprises four stages: starting with a batch size of $8$ configurations and $200$ epochs, with both the batch size and the number of epochs doubling at each subsequent stage. 

For each classes of molecular pair in the dimer tests, we employ a hybrid SR descriptor, including both $\bm{D}_{i,c}$ and $\bm{D}_{i,s}$, in our SOG-Net model. The cutoff radii and scaling factors are set to $10\mathring{A}$ and $3/4$ for the both components. In our model, the sizes of the embedding networks $f_{\bm{\theta}_{emb,c}}$ and $f_{\bm{\theta}_{emb,s}}$ are configured as $(20,40,80)$, and the parameter $M_c^<$ for constructing $\bm{D}_{i,c}$ is set as $80$. The sizes of the fitting network $f_{\bm{\theta}_{fit}}$ and the latent network $f_{\bm{\theta}_{lnn}}$ are set to $(64,64,64)$ and $(20,40,80,1)$, respectively. Consequently, a single latent variable layer ($P=1$) is sufficient for this set of tests. All networks uses the tanh activation function after each layer. Parameters for the LR component of the SOG-Net model are detailed in the main text. The models are trained using the Adam optimizer with an initial learning rate of $0.001$, which decays by a factor of $0.98$ every $10$ epochs. The training process comprises of three stages: starting with a batch size of $2$ configurations and $800$ epochs, with both the batch size and the number of epochs doubling at each subsequent stage. Note that we compare SOG-Net with the latent Ewald summation (LES) method ~\cite{cheng2024latent} in the main text. The LES results, taken from the original paper, were obtained using a CACE\cite{cheng2024cartesian} SR descriptor with a cutoff of $r_c=5\mathring{\text{A}}$ and one message-passing layer, giving an effective receptive field of approximately $10\mathring{\text{A}}$. Accordingly, both the CACE and DeepMD descriptors used in our tests capture a similar local environment. These comparisons are intended to demonstrate SOG-Net’s ability to capture diverse LR decay behaviors, though the results may also reflect differences in training protocols across works.

For the liquid water tests, we again use the core-shell structured SR descriptor in our SOG-Net model, with cutoff radii set to $4\mathring{A}$ and $7\mathring{A}$ for the core and shell components, respectively. The dimensions of embedding matrices and fitting matrix are set to $(25, 50, 100)$ and $(120, 120, 120)$, respectively. The scaling factors $j_c$ and $j_s$ used in the core and shell components are set to $1/8$ and $1/14$. The parameter $M^{<}_{c}$ for the core component is set as $16$. The sizes of the latent network
are configured as $(25, 50, 100, 16)$, so that the number of layer of latent variables is set as $16$. All networks uses the tanh activation function after each layer. Parameters for the LR component of the SOG-Net model are detailed in the main text. The models are trained using the Adam optimizer with an initial learning rate of $0.001$, which decays by a factor of $0.997$ every $10$ epochs. The training process comprises of two stages: starting with a batch size of $50$ configurations and $10000$ epochs, with both the batch size and the number of epochs doubling at the subsequent stage. The MD simulations are performed in the ASE package, and the LAMMPS interface implementation is our future work. To compute the charge structure factors,
we are assuming that hydrogen atoms and oxygen atoms have charges of $+0.4238e$ and $-0.8476e$, respectively, similar to previous works~\cite{cheng2024latent}. This assumption only affects the absolute amplitude of the correlation function in Figure~4 of the main text, without altering the relative scale.

For the C$_{10}$H$_2$/C$_{10}$H$_3^+$, Na$_{8/9}$Cl$_{8}^{+}$, Au$_2$ clusters on MgO(001), polar dipeptide, and molten bulk sodium chloride tests, we use the CACE~\cite{cheng2024cartesian} SR descriptor in our SOG-Net model. Comparison results, other than the SOG-Net model, for the first four datasets and the final dataset were obtained from existing studies~\cite{kim2024learning,cheng2024latent}, respectively. Accordingly, we set the cutoff radii to $r_c=4.23$, $5.29$, $5.5$, $4.0$, and $6.0~\mathring{\text{A}}$, respectively, to match the settings in these references. Other parameters of SR descriptor are set to the same as in~\cite{kim2024learning}: For C$_{10}$H$_2$/C$_{10}$H$_3^+$, we use $6$ Bessel radial functions, $c=8$, $l_{\text{max}}=3$, $\nu_{\text{max}}=3$, $N_{\text{embedding}}=2$, and no message passing. For Na$_{8/9}$Cl$_{8}^{+}$, we use $6$ Bessel radial functions, $c=8$, $l_{\text{max}}=3$, $\nu_{\text{max}}=3$, $N_{\text{embedding}}=2$, and no message passing. For Au$_2$ clusters on MgO(001), we use $6$ Bessel radial functions, $c=12$, $l_{\text{max}}=3$, $\nu_{\text{max}}=3$, $N_{\text{embedding}}=4$, and no message passing. For polar dipeptide tests, we use $6$ trainable Bessel radial functions, $c=12$, $l_{\text{max}}=4$, $\nu_{\text{max}}=3$, one message passing layer, and different embeddings of sender and receiver nodes with $N_{\text{embedding}}=4$. For SOG-Net, the LR energy from Eq.~\eqref{eq::phiexpan} is computed in real space as the configurations are with aperiodic conditions. The latent variables ($P=1$) are interpreted as atomic SOG-Net charges and used to compute dipole and quadrupole moments. To ensure translational invariance in quadrupole comparison, we subtract the trace from both the predicted and DFT quadrupole moments. The molten sodium chloride dataset contains 1014 structures (80$\%$ train and $20\%$ validation) of $64$ Na and 64 Cl atoms. The
 configurations are taken from VASP MD simulations with 50000 steps and a time step of $1.5$ fs. The PAW potentials used were ``PAW$\_$PBE
 Na$\_$pv 19Sep2006'' for Na and ``PAW$\_$PBE Cl 06Sep2006'',  for Cl, considering seven valence electrons for each element. We use $6$ Bessel radial functions, $c=12$, $l_{\text{max}}=3$, $\nu_{\text{max}}=3$, $N_{\text{embedding}}=3$, and zero or one message passing layer. For the SOG-Net, we use a $4$-layer latent variable, $M=12$ Gaussians, and $31$ Fourier modes along each direction.

\section{Supplementary results}\label{supplementary results}

\begin{figure*}[ht] 
    \centering
    \includegraphics[width=0.6\textwidth]{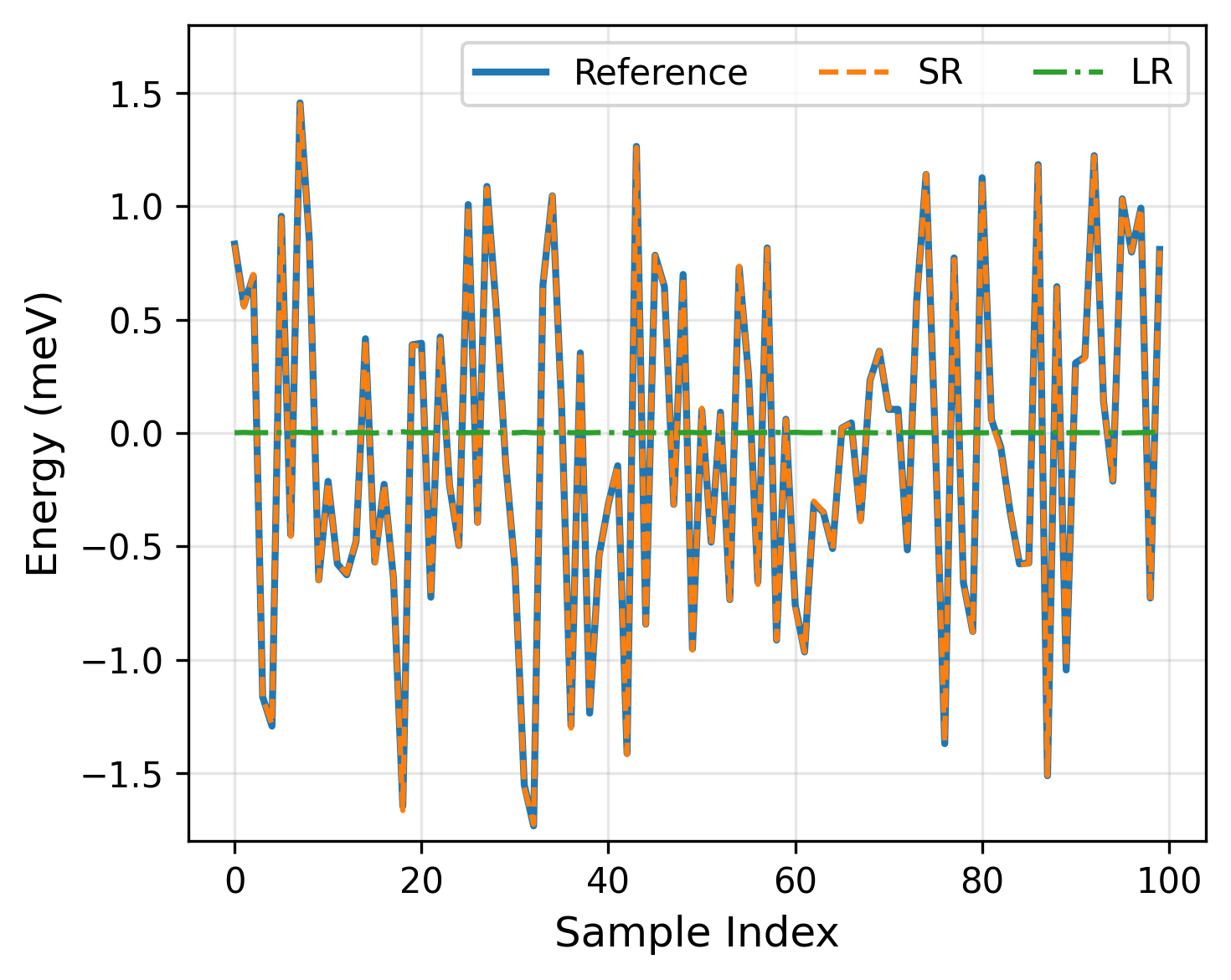}
    \caption{We tested SOG-Net on a purely short-range (SR) dataset, reporting the reference solution alongside the SR and long-range (LR) components predicted by the model. The relative $L_2$-norm of the LR energy contribution with respect to the total energy (SR + LR)  across all samples is approximately $1 \times 10^{-3}$, indicating that the LR component in SOG-Net is negligible for this dataset.}
    \label{fig:Pure}
\end{figure*}

\begin{table}[ht]
\renewcommand\arraystretch{1.2}
\centering
\begin{tabular}{|l|c|c|}
\hline & RMSPE E (\%) & RMSPE F (\%) \\
\hline SOAP\cite{bartok2010gaussian}  & 7.7 & 12.6 \\
\hline MACE\cite{batatia2022mace}  $\mathrm{~T}=0$ & 8.5 & 8.5 \\
\hline MACE\cite{batatia2022mace}  $\mathrm{~T}=1$ invariant & 3.0 & 3.1 \\
\hline MACE\cite{batatia2022mace} $\mathrm{~T}=1$ equivariant & 2.1 & 2.2 \\
\hline CACE\cite{cheng2024cartesian} $~\mathrm{T}=0$ & 4.8 & 8.2 \\
\hline CACE\cite{cheng2024cartesian} + fixed q = 1e & 7.3 & 10.5 \\
\hline CACE\cite{cheng2024cartesian} + fixed q = 0.65e & 3.0 & 4.3 \\
\hline LODE minimal\cite{Grisafi2019JCP} & 5.6 & 3.7 \\
\hline LODE flexible\cite{Grisafi2019JCP} & 7.0 & 11.0 \\
\hline Density-LR\cite{faller2024density} & 2.2 & 3.3 \\
\hline CACE-LR\cite{cheng2024latent} $\mathrm{~T}=0$ & 1.9 & 2.3 \\
\hline CACE-LR\cite{cheng2024latent} $\mathrm{~T}=1$ & 1.4 & 1.8 \\
\hline SOG-Net $\mathrm{~T}=0$ & 1.7 & 2.3\\
\hline SOG-Net $\mathrm{~T}=1$ & 1.4 & 1.9\\
\hline
\end{tabular}

\caption{The root-mean-square percentage error (RMSPE) of energy and force for the validation set of molten bulk sodium chloride system. This dataset is generated using PAW$\_$PBE (Na$\_$pv$/$Cl) pseudopotentials and features a long-range decay of $1/r$. Data are shown for the SOG-Net method and comparisons including the short-ranged SOAP \cite{bartok2010gaussian} descriptor, two setting of LODE \cite{Grisafi2019JCP} descriptors (minimal and flexible), MACE  \cite{batatia2022mace} with zero (T=0) or one message passing layers (T=1) as well as using in variant and equivariant messages, a density-based long-range model (Density-LR\cite{faller2024density}), and CACE-LR \cite{cheng2024latent} with zero (T=0) or one message passing layers (T=1). SOG-Net achieves performance comparable to the corresponding Ewald-based CACE-LR result, both with and without message-passing layers, and outperforms other models.}
\label{tab::TableComparison}
\end{table}

\begin{figure*}[ht] 
    \centering
    \includegraphics[width=0.63\textwidth]{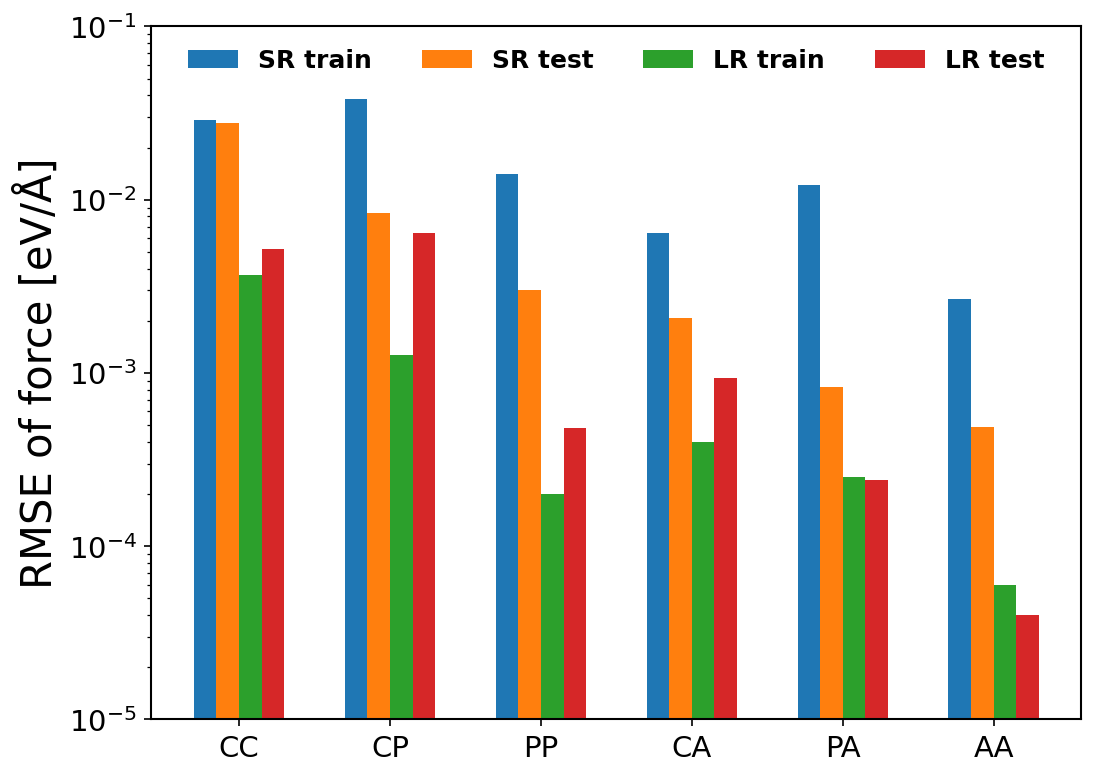}
    \caption{Root mean square errors (RMSE) of forces for the test of dimer pairs. We report the training and testing errors using pure short-range (SR) and full SOG-Net (LR) models. For the SOG-Net model, the number of Fourier grids is set to $31$ along each direction. Results are provided for charge-charge (CC), charge-polar (CP), polar-polar (PP), charge-apolar (CA), polar-apolar (PA), and apolar-apolar (AA) pairs.}
    \label{fig:SOG_Bar}
\end{figure*}



\begin{figure*}[ht] 
    \centering
    \includegraphics[width=0.6971\textwidth]{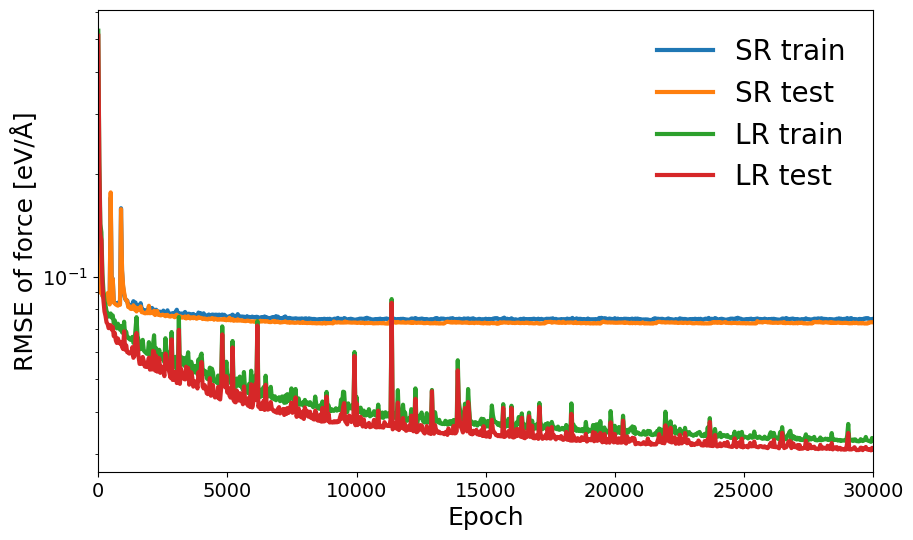}
    \caption{Root-mean-square errors (RMSEs) of forces during the training process for liquid water tests. Results are presented for both the pure short-range (SR) component and the full SOG-Net model, evaluated on training and test datasets.}
\label{fig:SOG_Loss}
\end{figure*}

\begin{figure*}[ht] 
    \centering
    \includegraphics[width=0.9\textwidth]{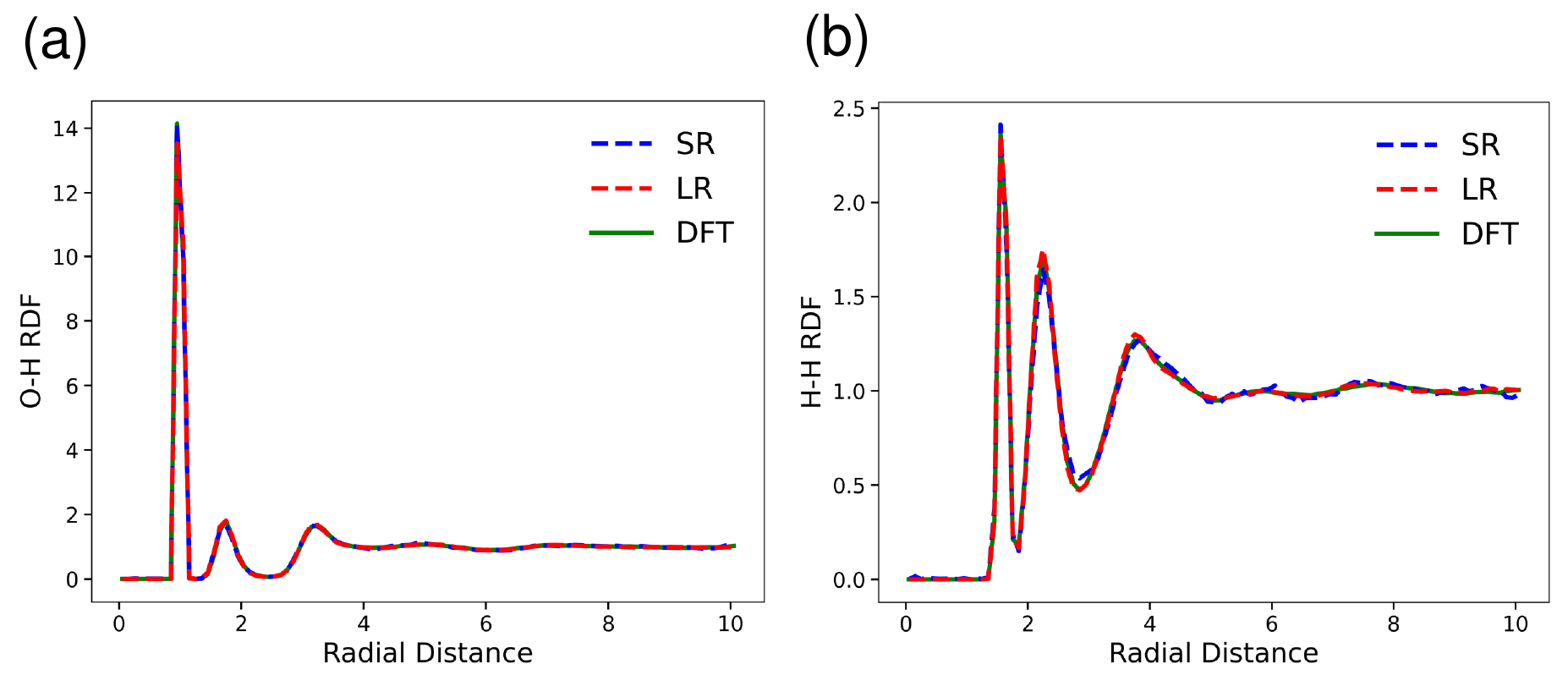}
    \caption{Predicted radial distribution functions (RDFs) of (a) oxygen-hydrogen (O-H) and (b) hydrogen-hydrogen (H-H) atom pairs of water at $300 K$ and $1 g/mL$, using short-range (SR) or long-range (LR) models compared with the DFT results.}
    \label{fig:RDF_rdf}
\end{figure*}

\begin{figure}[ht]
    \centering  \includegraphics[width=0.60\linewidth]{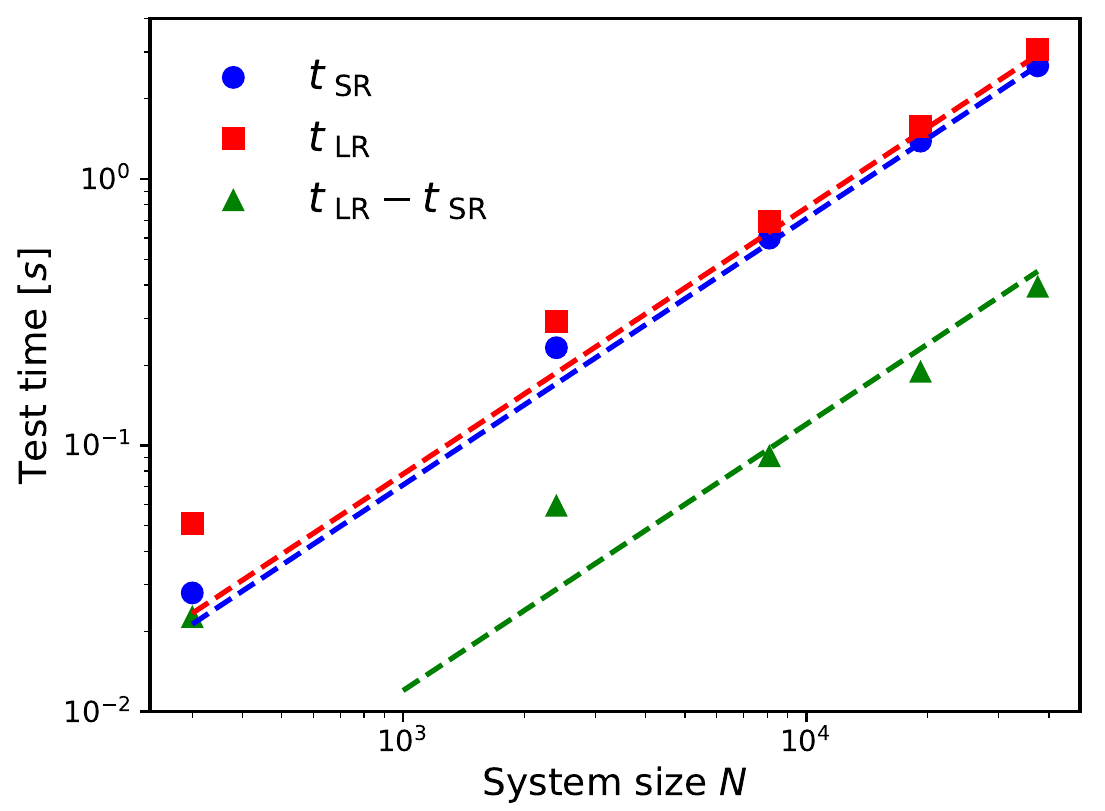}
    \caption{CPU time per step for the SOG-Net framework with increasing system size $N$ while simulating liquid water systems. Data are shown for timing results obtained using the pure SR network ($t_{\text{SR}}$), the full SOG-Net ($t_{\text{LR}}$), and the additional cost beyond the pure SR model ($t_{\text{LR}}-t_{\text{SR}}$). The dashed lines show the linear fitting of data.}
    \label{fig:enter-label}
\end{figure}



\begin{table}
\centering
\begin{tabular}{|c|c|c|c|c|c|c|}
\hline
\multirow{2}{*}{}
& CACE-SR & CACE-LR & SOG-Net & CACE-SR & CACE-LR& SOG-Net\\
& Val & Val & Val & Test & Test &Test\\
\hline E & 1.72 & 1.45 & \textbf{0.79}& 2.35 & 1.89 & \textbf{1.28} \\
\hline F & 58.07 & 52.18 & \textbf{49.52} & 72.43 & \textbf{61.13}& 61.15 \\
\hline
\end{tabular}
    \caption{Performance of CACE-SR\cite{cheng2024cartesian}, CACE-LR\cite{cheng2024latent}, and SOG-Net on the validation and test sets of the polar dipeptide dataset. Errors are reported via RMSE in meV/atom for energy and in $\mathrm{meV} / \mathring{\text{A}}$ for forces.}
    \label{tab:my_labelAccuracy}
\end{table}

\begin{figure}[ht]
\centering
\includegraphics[width=0.35\linewidth]{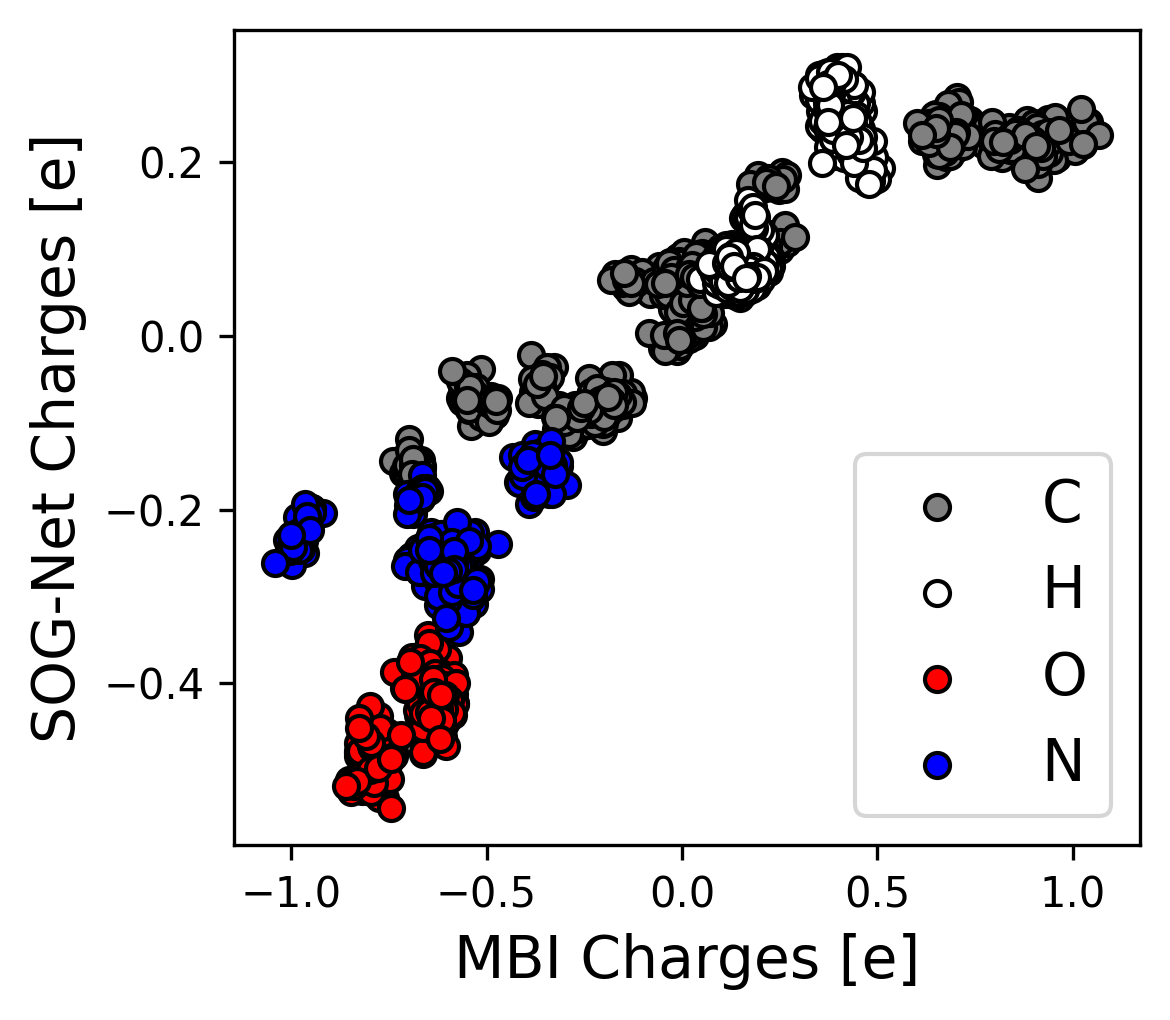}
\includegraphics[width=0.31\linewidth]{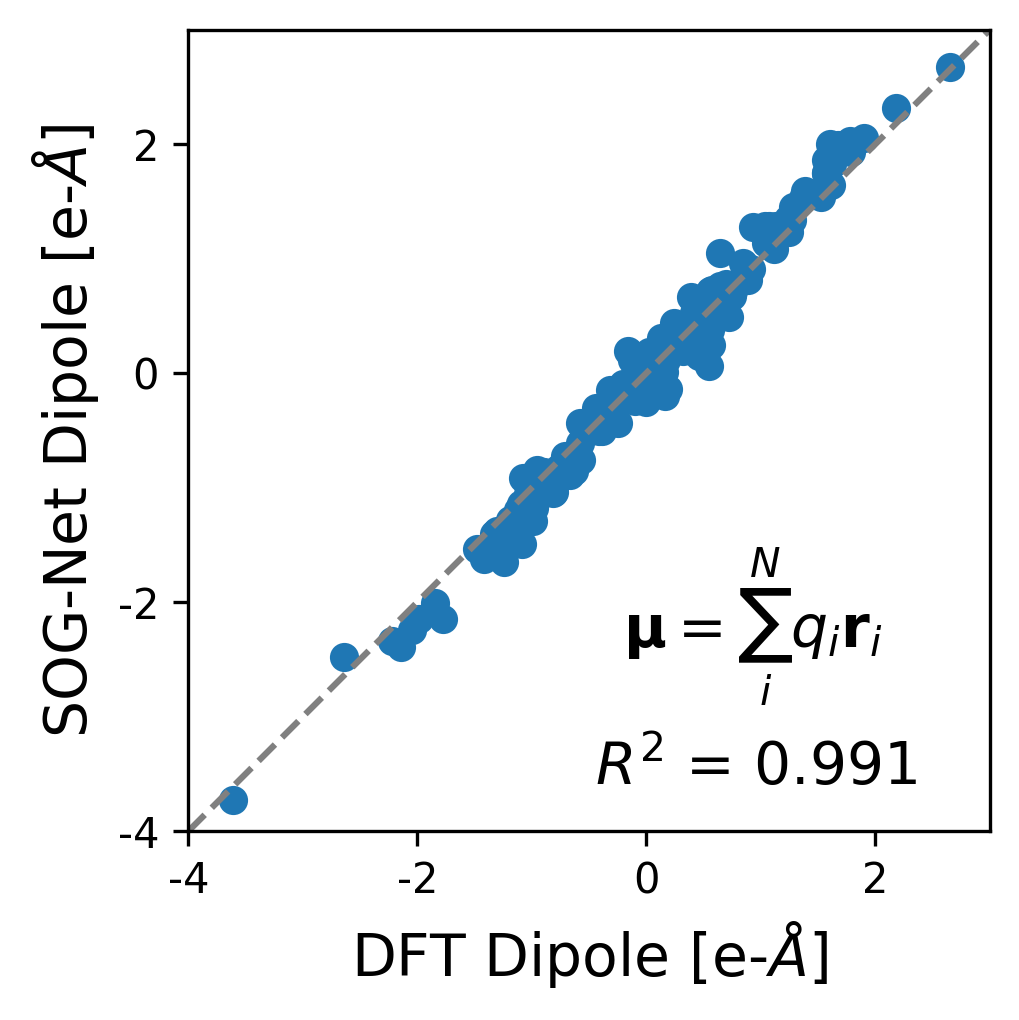}
\includegraphics[width=0.32\linewidth]{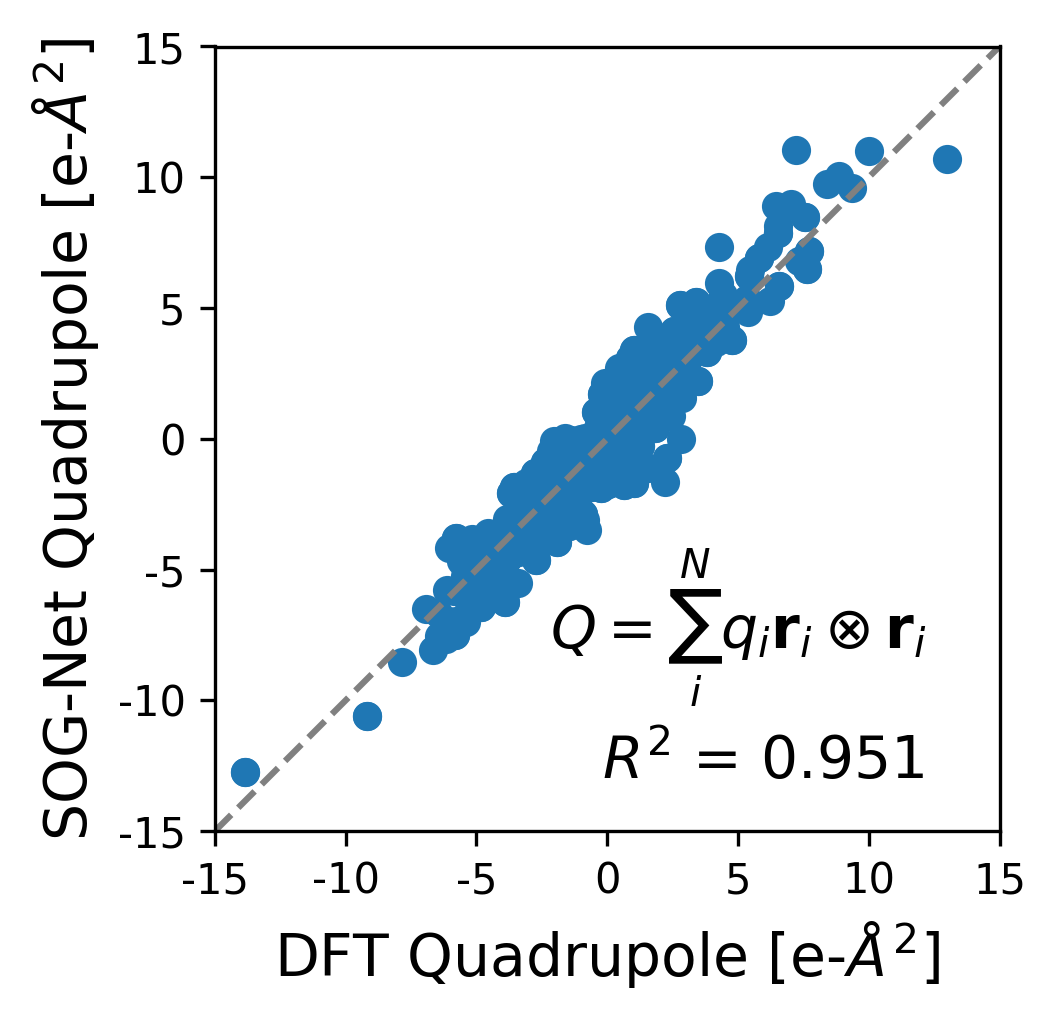}
\caption{Tests on polar dipeptides. (a) The predicted charges from SOG-Net (latent variables) compared to minimal basis iterative stockholder (MBI) charges in SPICE. They correlate well and follow the expected electronegativity trend (O$>$N$>$C$>$H). (b) The predicted dipole components computed from the SOG-Net charges ($\mu=\sum^N_{i=1}q_ir_i$) compared to the DFT dipole components in SPICE. (c) The predicted traceless quadrupole components computed from the SOG-Net charges ($Q=\sum^N_{
i=1}q_ir_i\otimes r_i$) compared to the DFT quadrupole components in SPICE.}
\label{fig:LRtransfer}
\end{figure}


\begin{thebibliography}{61}%
\makeatletter
\providecommand \@ifxundefined [1]{%
 \@ifx{#1\undefined}
}%
\providecommand \@ifnum [1]{%
 \ifnum #1\expandafter \@firstoftwo
 \else \expandafter \@secondoftwo
 \fi
}%
\providecommand \@ifx [1]{%
 \ifx #1\expandafter \@firstoftwo
 \else \expandafter \@secondoftwo
 \fi
}%
\providecommand \natexlab [1]{#1}%
\providecommand \enquote  [1]{``#1''}%
\providecommand \bibnamefont  [1]{#1}%
\providecommand \bibfnamefont [1]{#1}%
\providecommand \citenamefont [1]{#1}%
\providecommand \href@noop [0]{\@secondoftwo}%
\providecommand \href [0]{\begingroup \@sanitize@url \@href}%
\providecommand \@href[1]{\@@startlink{#1}\@@href}%
\providecommand \@@href[1]{\endgroup#1\@@endlink}%
\providecommand \@sanitize@url [0]{\catcode `\\12\catcode `\$12\catcode `\&12\catcode `\#12\catcode `\^12\catcode `\_12\catcode `\%12\relax}%
\providecommand \@@startlink[1]{}%
\providecommand \@@endlink[0]{}%
\providecommand \url  [0]{\begingroup\@sanitize@url \@url }%
\providecommand \@url [1]{\endgroup\@href {#1}{\urlprefix }}%
\providecommand \urlprefix  [0]{URL }%
\providecommand \Eprint [0]{\href }%
\providecommand \doibase [0]{https://doi.org/}%
\providecommand \selectlanguage [0]{\@gobble}%
\providecommand \bibinfo  [0]{\@secondoftwo}%
\providecommand \bibfield  [0]{\@secondoftwo}%
\providecommand \translation [1]{[#1]}%
\providecommand \BibitemOpen [0]{}%
\providecommand \bibitemStop [0]{}%
\providecommand \bibitemNoStop [0]{.\EOS\space}%
\providecommand \EOS [0]{\spacefactor3000\relax}%
\providecommand \BibitemShut  [1]{\csname bibitem#1\endcsname}%
\let\auto@bib@innerbib\@empty
\bibitem [{\citenamefont {Friederich}\ \emph {et~al.}(2021)\citenamefont {Friederich}, \citenamefont {H{\"a}se}, \citenamefont {Proppe},\ and\ \citenamefont {Aspuru-Guzik}}]{friederich2021machine}%
  \BibitemOpen
  \bibfield  {author} {\bibinfo {author} {\bibfnamefont {P.}~\bibnamefont {Friederich}}, \bibinfo {author} {\bibfnamefont {F.}~\bibnamefont {H{\"a}se}}, \bibinfo {author} {\bibfnamefont {J.}~\bibnamefont {Proppe}},\ and\ \bibinfo {author} {\bibfnamefont {A.}~\bibnamefont {Aspuru-Guzik}},\ }\bibfield  {title} {\bibinfo {title} {Machine-learned potentials for next-generation matter simulations},\ }\href@noop {} {\bibfield  {journal} {\bibinfo  {journal} {Nat. Mater.}\ }\textbf {\bibinfo {volume} {20}},\ \bibinfo {pages} {750} (\bibinfo {year} {2021})}\BibitemShut {NoStop}%
\bibitem [{\citenamefont {Unke}\ \emph {et~al.}(2021{\natexlab{a}})\citenamefont {Unke}, \citenamefont {Chmiela}, \citenamefont {Sauceda}, \citenamefont {Gastegger}, \citenamefont {Poltavsky}, \citenamefont {Sch{\"u}tt}, \citenamefont {Tkatchenko},\ and\ \citenamefont {M{\"u}ller}}]{unke2021machine}%
  \BibitemOpen
  \bibfield  {author} {\bibinfo {author} {\bibfnamefont {O.~T.}\ \bibnamefont {Unke}}, \bibinfo {author} {\bibfnamefont {S.}~\bibnamefont {Chmiela}}, \bibinfo {author} {\bibfnamefont {H.~E.}\ \bibnamefont {Sauceda}}, \bibinfo {author} {\bibfnamefont {M.}~\bibnamefont {Gastegger}}, \bibinfo {author} {\bibfnamefont {I.}~\bibnamefont {Poltavsky}}, \bibinfo {author} {\bibfnamefont {K.~T.}\ \bibnamefont {Sch{\"u}tt}}, \bibinfo {author} {\bibfnamefont {A.}~\bibnamefont {Tkatchenko}},\ and\ \bibinfo {author} {\bibfnamefont {K.~R.}\ \bibnamefont {M{\"u}ller}},\ }\bibfield  {title} {\bibinfo {title} {Machine learning force fields},\ }\href@noop {} {\bibfield  {journal} {\bibinfo  {journal} {Chem. Rev.}\ }\textbf {\bibinfo {volume} {121}},\ \bibinfo {pages} {10142} (\bibinfo {year} {2021}{\natexlab{a}})}\BibitemShut {NoStop}%
\bibitem [{\citenamefont {Zhang}\ \emph {et~al.}(2018)\citenamefont {Zhang}, \citenamefont {Han}, \citenamefont {Wang}, \citenamefont {Car},\ and\ \citenamefont {E}}]{zhang2018deep}%
  \BibitemOpen
  \bibfield  {author} {\bibinfo {author} {\bibfnamefont {L.}~\bibnamefont {Zhang}}, \bibinfo {author} {\bibfnamefont {J.}~\bibnamefont {Han}}, \bibinfo {author} {\bibfnamefont {H.}~\bibnamefont {Wang}}, \bibinfo {author} {\bibfnamefont {R.}~\bibnamefont {Car}},\ and\ \bibinfo {author} {\bibfnamefont {W.}~\bibnamefont {E}},\ }\bibfield  {title} {\bibinfo {title} {Deep potential molecular dynamics: a scalable model with the accuracy of quantum mechanics},\ }\href@noop {} {\bibfield  {journal} {\bibinfo  {journal} {Phys. Rev. Lett.}\ }\textbf {\bibinfo {volume} {120}},\ \bibinfo {pages} {143001} (\bibinfo {year} {2018})}\BibitemShut {NoStop}%
\bibitem [{\citenamefont {Kohn}(1996)}]{kohn1996density}%
  \BibitemOpen
  \bibfield  {author} {\bibinfo {author} {\bibfnamefont {W.}~\bibnamefont {Kohn}},\ }\bibfield  {title} {\bibinfo {title} {Density functional and density matrix method scaling linearly with the number of atoms},\ }\href@noop {} {\bibfield  {journal} {\bibinfo  {journal} {Phys. Rev. Lett.}\ }\textbf {\bibinfo {volume} {76}},\ \bibinfo {pages} {3168} (\bibinfo {year} {1996})}\BibitemShut {NoStop}%
\bibitem [{\citenamefont {Prodan}\ and\ \citenamefont {Kohn}(2005)}]{PNAS2005}%
  \BibitemOpen
  \bibfield  {author} {\bibinfo {author} {\bibfnamefont {E.}~\bibnamefont {Prodan}}\ and\ \bibinfo {author} {\bibfnamefont {W.}~\bibnamefont {Kohn}},\ }\bibfield  {title} {\bibinfo {title} {Nearsightedness of electronic matter},\ }\href@noop {} {\bibfield  {journal} {\bibinfo  {journal} {P. Nat. Acad. Sci.}\ }\textbf {\bibinfo {volume} {102}},\ \bibinfo {pages} {11635} (\bibinfo {year} {2005})}\BibitemShut {NoStop}%
\bibitem [{\citenamefont {Zhou}\ and\ \citenamefont {Pang}(2018)}]{zhou2018electrostatic}%
  \BibitemOpen
  \bibfield  {author} {\bibinfo {author} {\bibfnamefont {H.-X.}\ \bibnamefont {Zhou}}\ and\ \bibinfo {author} {\bibfnamefont {X.}~\bibnamefont {Pang}},\ }\bibfield  {title} {\bibinfo {title} {Electrostatic interactions in protein structure, folding, binding, and condensation},\ }\href@noop {} {\bibfield  {journal} {\bibinfo  {journal} {Chem. Rev.}\ }\textbf {\bibinfo {volume} {118}},\ \bibinfo {pages} {1691} (\bibinfo {year} {2018})}\BibitemShut {NoStop}%
\bibitem [{\citenamefont {Yuan}\ and\ \citenamefont {Tanaka}(2024)}]{PhysRevLett.132.228101}%
  \BibitemOpen
  \bibfield  {author} {\bibinfo {author} {\bibfnamefont {J.}~\bibnamefont {Yuan}}\ and\ \bibinfo {author} {\bibfnamefont {H.}~\bibnamefont {Tanaka}},\ }\bibfield  {title} {\bibinfo {title} {Charge regulation effects in polyelectrolyte adsorption},\ }\href@noop {} {\bibfield  {journal} {\bibinfo  {journal} {Phys. Rev. Lett.}\ }\textbf {\bibinfo {volume} {132}},\ \bibinfo {pages} {228101} (\bibinfo {year} {2024})}\BibitemShut {NoStop}%
\bibitem [{\citenamefont {Bart{\'o}k}\ \emph {et~al.}(2010)\citenamefont {Bart{\'o}k}, \citenamefont {Payne}, \citenamefont {Kondor},\ and\ \citenamefont {Cs{\'a}nyi}}]{bartok2010gaussian}%
  \BibitemOpen
  \bibfield  {author} {\bibinfo {author} {\bibfnamefont {A.~P.}\ \bibnamefont {Bart{\'o}k}}, \bibinfo {author} {\bibfnamefont {M.~C.}\ \bibnamefont {Payne}}, \bibinfo {author} {\bibfnamefont {R.}~\bibnamefont {Kondor}},\ and\ \bibinfo {author} {\bibfnamefont {G.}~\bibnamefont {Cs{\'a}nyi}},\ }\bibfield  {title} {\bibinfo {title} {Gaussian approximation potentials: {T}he accuracy of quantum mechanics, without the electrons},\ }\href@noop {} {\bibfield  {journal} {\bibinfo  {journal} {Phys. Rev. Lett.}\ }\textbf {\bibinfo {volume} {104}},\ \bibinfo {pages} {136403} (\bibinfo {year} {2010})}\BibitemShut {NoStop}%
\bibitem [{\citenamefont {Unke}\ \emph {et~al.}(2021{\natexlab{b}})\citenamefont {Unke}, \citenamefont {Chmiela}, \citenamefont {Gastegger}, \citenamefont {Sch{\"u}tt}, \citenamefont {Sauceda},\ and\ \citenamefont {M{\"u}ller}}]{Unke2021}%
  \BibitemOpen
  \bibfield  {author} {\bibinfo {author} {\bibfnamefont {O.~T.}\ \bibnamefont {Unke}}, \bibinfo {author} {\bibfnamefont {S.}~\bibnamefont {Chmiela}}, \bibinfo {author} {\bibfnamefont {M.}~\bibnamefont {Gastegger}}, \bibinfo {author} {\bibfnamefont {K.~T.}\ \bibnamefont {Sch{\"u}tt}}, \bibinfo {author} {\bibfnamefont {H.~E.}\ \bibnamefont {Sauceda}},\ and\ \bibinfo {author} {\bibfnamefont {K.~R.}\ \bibnamefont {M{\"u}ller}},\ }\bibfield  {title} {\bibinfo {title} {{SpookyNet: Learning force fields with electronic degrees of freedom and nonlocal effects}},\ }\href@noop {} {\bibfield  {journal} {\bibinfo  {journal} {Nat. Commun.}\ }\textbf {\bibinfo {volume} {12}},\ \bibinfo {pages} {7273} (\bibinfo {year} {2021}{\natexlab{b}})}\BibitemShut {NoStop}%
\bibitem [{\citenamefont {Zhang}\ \emph {et~al.}(2022)\citenamefont {Zhang}, \citenamefont {Wang}, \citenamefont {Muniz}, \citenamefont {Panagiotopoulos}, \citenamefont {Car} \emph {et~al.}}]{zhang2022deep}%
  \BibitemOpen
  \bibfield  {author} {\bibinfo {author} {\bibfnamefont {L.}~\bibnamefont {Zhang}}, \bibinfo {author} {\bibfnamefont {H.}~\bibnamefont {Wang}}, \bibinfo {author} {\bibfnamefont {M.~C.}\ \bibnamefont {Muniz}}, \bibinfo {author} {\bibfnamefont {A.~Z.}\ \bibnamefont {Panagiotopoulos}}, \bibinfo {author} {\bibfnamefont {R.}~\bibnamefont {Car}}, \emph {et~al.},\ }\bibfield  {title} {\bibinfo {title} {A deep potential model with long-range electrostatic interactions},\ }\href@noop {} {\bibfield  {journal} {\bibinfo  {journal} {J. Chem. Phys.}\ }\textbf {\bibinfo {volume} {156}},\ \bibinfo {pages} {124107} (\bibinfo {year} {2022})}\BibitemShut {NoStop}%
\bibitem [{\citenamefont {Gao}\ and\ \citenamefont {Remsing}(2022)}]{gao2022self}%
  \BibitemOpen
  \bibfield  {author} {\bibinfo {author} {\bibfnamefont {A.}~\bibnamefont {Gao}}\ and\ \bibinfo {author} {\bibfnamefont {R.~C.}\ \bibnamefont {Remsing}},\ }\bibfield  {title} {\bibinfo {title} {Self-consistent determination of long-range electrostatics in neural network potentials},\ }\href@noop {} {\bibfield  {journal} {\bibinfo  {journal} {Nat. Commun.}\ }\textbf {\bibinfo {volume} {13}},\ \bibinfo {pages} {1572} (\bibinfo {year} {2022})}\BibitemShut {NoStop}%
\bibitem [{\citenamefont {Unke}\ and\ \citenamefont {Meuwly}(2019)}]{unke2019physnet}%
  \BibitemOpen
  \bibfield  {author} {\bibinfo {author} {\bibfnamefont {O.~T.}\ \bibnamefont {Unke}}\ and\ \bibinfo {author} {\bibfnamefont {M.}~\bibnamefont {Meuwly}},\ }\bibfield  {title} {\bibinfo {title} {{PhysNet: A neural network for predicting energies, forces, dipole moments, and partial charges}},\ }\href@noop {} {\bibfield  {journal} {\bibinfo  {journal} {J. Chem. Theory Comput.}\ }\textbf {\bibinfo {volume} {15}},\ \bibinfo {pages} {3678} (\bibinfo {year} {2019})}\BibitemShut {NoStop}%
\bibitem [{\citenamefont {Ko}\ \emph {et~al.}(2021)\citenamefont {Ko}, \citenamefont {Finkler}, \citenamefont {Goedecker},\ and\ \citenamefont {Behler}}]{ko2021fourth}%
  \BibitemOpen
  \bibfield  {author} {\bibinfo {author} {\bibfnamefont {T.~W.}\ \bibnamefont {Ko}}, \bibinfo {author} {\bibfnamefont {J.~A.}\ \bibnamefont {Finkler}}, \bibinfo {author} {\bibfnamefont {S.}~\bibnamefont {Goedecker}},\ and\ \bibinfo {author} {\bibfnamefont {J.}~\bibnamefont {Behler}},\ }\bibfield  {title} {\bibinfo {title} {A fourth-generation high-dimensional neural network potential with accurate electrostatics including non-local charge transfer},\ }\href@noop {} {\bibfield  {journal} {\bibinfo  {journal} {Nat. Commun.}\ }\textbf {\bibinfo {volume} {12}},\ \bibinfo {pages} {398} (\bibinfo {year} {2021})}\BibitemShut {NoStop}%
\bibitem [{\citenamefont {Shaidu}\ \emph {et~al.}(2024)\citenamefont {Shaidu}, \citenamefont {Pellegrini}, \citenamefont {K{\"u}{c}{\"u}kbenli}, \citenamefont {Lot},\ and\ \citenamefont {de~Gironcoli}}]{shaidu2024incorporating}%
  \BibitemOpen
  \bibfield  {author} {\bibinfo {author} {\bibfnamefont {Y.}~\bibnamefont {Shaidu}}, \bibinfo {author} {\bibfnamefont {F.}~\bibnamefont {Pellegrini}}, \bibinfo {author} {\bibfnamefont {E.}~\bibnamefont {K{\"u}{c}{\"u}kbenli}}, \bibinfo {author} {\bibfnamefont {R.}~\bibnamefont {Lot}},\ and\ \bibinfo {author} {\bibfnamefont {S.}~\bibnamefont {de~Gironcoli}},\ }\bibfield  {title} {\bibinfo {title} {Incorporating long-range electrostatics in neural network potentials via variational charge equilibration from shortsighted ingredients},\ }\href@noop {} {\bibfield  {journal} {\bibinfo  {journal} {npj Comput. Mater.}\ }\textbf {\bibinfo {volume} {10}},\ \bibinfo {pages} {47} (\bibinfo {year} {2024})}\BibitemShut {NoStop}%
\bibitem [{\citenamefont {Ewald}(1921)}]{ewald1921berechnung}%
  \BibitemOpen
  \bibfield  {author} {\bibinfo {author} {\bibfnamefont {P.~P.}\ \bibnamefont {Ewald}},\ }\bibfield  {title} {\bibinfo {title} {{Die Berechnung optischer und elektrostatischer Gitterpotentiale}},\ }\href@noop {} {\bibfield  {journal} {\bibinfo  {journal} {Ann. Phys.}\ }\textbf {\bibinfo {volume} {369}},\ \bibinfo {pages} {253} (\bibinfo {year} {1921})}\BibitemShut {NoStop}%
\bibitem [{\citenamefont {Zhang}\ \emph {et~al.}(2021)\citenamefont {Zhang}, \citenamefont {Wang}, \citenamefont {Car},\ and\ \citenamefont {E}}]{zhang2021phase}%
  \BibitemOpen
  \bibfield  {author} {\bibinfo {author} {\bibfnamefont {L.}~\bibnamefont {Zhang}}, \bibinfo {author} {\bibfnamefont {H.}~\bibnamefont {Wang}}, \bibinfo {author} {\bibfnamefont {R.}~\bibnamefont {Car}},\ and\ \bibinfo {author} {\bibfnamefont {W.}~\bibnamefont {E}},\ }\bibfield  {title} {\bibinfo {title} {Phase diagram of a deep potential water model},\ }\href@noop {} {\bibfield  {journal} {\bibinfo  {journal} {Phys. Rev. Lett.}\ }\textbf {\bibinfo {volume} {126}},\ \bibinfo {pages} {236001} (\bibinfo {year} {2021})}\BibitemShut {NoStop}%
\bibitem [{\citenamefont {Bore}\ and\ \citenamefont {Paesani}(2023)}]{bore2023realistic}%
  \BibitemOpen
  \bibfield  {author} {\bibinfo {author} {\bibfnamefont {S.~L.}\ \bibnamefont {Bore}}\ and\ \bibinfo {author} {\bibfnamefont {F.}~\bibnamefont {Paesani}},\ }\bibfield  {title} {\bibinfo {title} {Realistic phase diagram of water from ``first principles'' data-driven quantum simulations},\ }\href@noop {} {\bibfield  {journal} {\bibinfo  {journal} {Nat. Commun.}\ }\textbf {\bibinfo {volume} {14}},\ \bibinfo {pages} {3349} (\bibinfo {year} {2023})}\BibitemShut {NoStop}%
\bibitem [{\citenamefont {Israelachvili}(2011)}]{israelachvili2011intermolecular}%
  \BibitemOpen
  \bibfield  {author} {\bibinfo {author} {\bibfnamefont {J.~N.}\ \bibnamefont {Israelachvili}},\ }\href@noop {} {\emph {\bibinfo {title} {{Intermolecular and Surface Forces}}}}\ (\bibinfo  {publisher} {Academic Press},\ \bibinfo {year} {2011})\BibitemShut {NoStop}%
\bibitem [{\citenamefont {Huguenin-Dumittan}\ \emph {et~al.}(2023)\citenamefont {Huguenin-Dumittan}, \citenamefont {Loche}, \citenamefont {Haoran},\ and\ \citenamefont {Ceriotti}}]{huguenin2023physics}%
  \BibitemOpen
  \bibfield  {author} {\bibinfo {author} {\bibfnamefont {K.~K.}\ \bibnamefont {Huguenin-Dumittan}}, \bibinfo {author} {\bibfnamefont {P.}~\bibnamefont {Loche}}, \bibinfo {author} {\bibfnamefont {N.}~\bibnamefont {Haoran}},\ and\ \bibinfo {author} {\bibfnamefont {M.}~\bibnamefont {Ceriotti}},\ }\bibfield  {title} {\bibinfo {title} {Physics-inspired equivariant descriptors of nonbonded interactions},\ }\href@noop {} {\bibfield  {journal} {\bibinfo  {journal} {J. Phys. Chem. Lett.}\ }\textbf {\bibinfo {volume} {14}},\ \bibinfo {pages} {9612} (\bibinfo {year} {2023})}\BibitemShut {NoStop}%
\bibitem [{\citenamefont {Grisafi}\ and\ \citenamefont {Ceriotti}(2019)}]{Grisafi2019JCP}%
  \BibitemOpen
  \bibfield  {author} {\bibinfo {author} {\bibfnamefont {A.}~\bibnamefont {Grisafi}}\ and\ \bibinfo {author} {\bibfnamefont {M.}~\bibnamefont {Ceriotti}},\ }\bibfield  {title} {\bibinfo {title} {{Incorporating long-range physics in atomic-scale machine learning}},\ }\href@noop {} {\bibfield  {journal} {\bibinfo  {journal} {J. Chem. Phys.}\ }\textbf {\bibinfo {volume} {151}},\ \bibinfo {pages} {204105} (\bibinfo {year} {2019})}\BibitemShut {NoStop}%
\bibitem [{\citenamefont {Sch{\"u}tt}\ \emph {et~al.}(2017)\citenamefont {Sch{\"u}tt}, \citenamefont {Kindermans}, \citenamefont {Sauceda~Felix}, \citenamefont {Chmiela}, \citenamefont {Tkatchenko},\ and\ \citenamefont {M{\"u}ller}}]{schutt2017schnet}%
  \BibitemOpen
  \bibfield  {author} {\bibinfo {author} {\bibfnamefont {K.}~\bibnamefont {Sch{\"u}tt}}, \bibinfo {author} {\bibfnamefont {P.-J.}\ \bibnamefont {Kindermans}}, \bibinfo {author} {\bibfnamefont {H.~E.}\ \bibnamefont {Sauceda~Felix}}, \bibinfo {author} {\bibfnamefont {S.}~\bibnamefont {Chmiela}}, \bibinfo {author} {\bibfnamefont {A.}~\bibnamefont {Tkatchenko}},\ and\ \bibinfo {author} {\bibfnamefont {K.~R.}\ \bibnamefont {M{\"u}ller}},\ }\bibfield  {title} {\bibinfo {title} {Schnet: A continuous-filter convolutional neural network for modeling quantum interactions},\ }\href@noop {} {\bibfield  {journal} {\bibinfo  {journal} {Adv. Neural Inf. Process. Syst. (NeurIPS)}\ }\textbf {\bibinfo {volume} {30}},\ \bibinfo {pages} {992} (\bibinfo {year} {2017})}\BibitemShut {NoStop}%
\bibitem [{\citenamefont {Batzner}\ \emph {et~al.}(2022)\citenamefont {Batzner}, \citenamefont {Musaelian}, \citenamefont {Sun}, \citenamefont {Geiger}, \citenamefont {Mailoa}, \citenamefont {Kornbluth}, \citenamefont {Molinari}, \citenamefont {Smidt},\ and\ \citenamefont {Kozinsky}}]{batzner20223}%
  \BibitemOpen
  \bibfield  {author} {\bibinfo {author} {\bibfnamefont {S.}~\bibnamefont {Batzner}}, \bibinfo {author} {\bibfnamefont {A.}~\bibnamefont {Musaelian}}, \bibinfo {author} {\bibfnamefont {L.}~\bibnamefont {Sun}}, \bibinfo {author} {\bibfnamefont {M.}~\bibnamefont {Geiger}}, \bibinfo {author} {\bibfnamefont {J.~P.}\ \bibnamefont {Mailoa}}, \bibinfo {author} {\bibfnamefont {M.}~\bibnamefont {Kornbluth}}, \bibinfo {author} {\bibfnamefont {N.}~\bibnamefont {Molinari}}, \bibinfo {author} {\bibfnamefont {T.~E.}\ \bibnamefont {Smidt}},\ and\ \bibinfo {author} {\bibfnamefont {B.}~\bibnamefont {Kozinsky}},\ }\bibfield  {title} {\bibinfo {title} {E(3)-equivariant graph neural networks for data-efficient and accurate interatomic potentials},\ }\href@noop {} {\bibfield  {journal} {\bibinfo  {journal} {Nat. Commun.}\ }\textbf {\bibinfo {volume} {13}},\ \bibinfo {pages} {2453} (\bibinfo {year} {2022})}\BibitemShut {NoStop}%
\bibitem [{\citenamefont {Batatia}\ \emph {et~al.}(2022)\citenamefont {Batatia}, \citenamefont {Kovacs}, \citenamefont {Simm}, \citenamefont {Ortner},\ and\ \citenamefont {Cs{\'a}nyi}}]{batatia2022mace}%
  \BibitemOpen
  \bibfield  {author} {\bibinfo {author} {\bibfnamefont {I.}~\bibnamefont {Batatia}}, \bibinfo {author} {\bibfnamefont {D.~P.}\ \bibnamefont {Kovacs}}, \bibinfo {author} {\bibfnamefont {G.}~\bibnamefont {Simm}}, \bibinfo {author} {\bibfnamefont {C.}~\bibnamefont {Ortner}},\ and\ \bibinfo {author} {\bibfnamefont {G.}~\bibnamefont {Cs{\'a}nyi}},\ }\bibfield  {title} {\bibinfo {title} {{MACE: Higher order equivariant message passing neural networks for fast and accurate force fields}},\ }\href@noop {} {\bibfield  {journal} {\bibinfo  {journal} {Adv. Neural Inf. Process. Syst. (NeurIPS)}\ }\textbf {\bibinfo {volume} {35}},\ \bibinfo {pages} {11423} (\bibinfo {year} {2022})}\BibitemShut {NoStop}%
\bibitem [{\citenamefont {Kosmala}\ \emph {et~al.}(2023)\citenamefont {Kosmala}, \citenamefont {Gasteiger}, \citenamefont {Gao},\ and\ \citenamefont {G{\"u}nnemann}}]{kosmala2023ewald}%
  \BibitemOpen
  \bibfield  {author} {\bibinfo {author} {\bibfnamefont {A.}~\bibnamefont {Kosmala}}, \bibinfo {author} {\bibfnamefont {J.}~\bibnamefont {Gasteiger}}, \bibinfo {author} {\bibfnamefont {N.}~\bibnamefont {Gao}},\ and\ \bibinfo {author} {\bibfnamefont {S.}~\bibnamefont {G{\"u}nnemann}},\ }\bibfield  {title} {\bibinfo {title} {Ewald-based long-range message passing for molecular graphs},\ }in\ \href@noop {} {\emph {\bibinfo {booktitle} {International Conference on Machine Learning}}}\ (\bibinfo {organization} {ICML},\ \bibinfo {year} {2023})\ pp.\ \bibinfo {pages} {17544--17563}\BibitemShut {NoStop}%
\bibitem [{\citenamefont {Deng}\ \emph {et~al.}(2023)\citenamefont {Deng}, \citenamefont {Zhong}, \citenamefont {Jun}, \citenamefont {Riebesell}, \citenamefont {Han}, \citenamefont {Bartel},\ and\ \citenamefont {Ceder}}]{deng2023chgnet}%
  \BibitemOpen
  \bibfield  {author} {\bibinfo {author} {\bibfnamefont {B.}~\bibnamefont {Deng}}, \bibinfo {author} {\bibfnamefont {P.}~\bibnamefont {Zhong}}, \bibinfo {author} {\bibfnamefont {K.}~\bibnamefont {Jun}}, \bibinfo {author} {\bibfnamefont {J.}~\bibnamefont {Riebesell}}, \bibinfo {author} {\bibfnamefont {K.}~\bibnamefont {Han}}, \bibinfo {author} {\bibfnamefont {C.~J.}\ \bibnamefont {Bartel}},\ and\ \bibinfo {author} {\bibfnamefont {G.}~\bibnamefont {Ceder}},\ }\bibfield  {title} {\bibinfo {title} {{CHGNet} as a pretrained universal neural network potential for charge-informed atomistic modelling},\ }\href@noop {} {\bibfield  {journal} {\bibinfo  {journal} {Nat. Mach. Intell.}\ }\textbf {\bibinfo {volume} {5}},\ \bibinfo {pages} {1031} (\bibinfo {year} {2023})}\BibitemShut {NoStop}%
\bibitem [{\citenamefont {Dutt}\ and\ \citenamefont {Rokhlin}(1993)}]{dutt1993fast}%
  \BibitemOpen
  \bibfield  {author} {\bibinfo {author} {\bibfnamefont {A.}~\bibnamefont {Dutt}}\ and\ \bibinfo {author} {\bibfnamefont {V.}~\bibnamefont {Rokhlin}},\ }\bibfield  {title} {\bibinfo {title} {Fast {F}ourier transforms for nonequispaced data},\ }\href@noop {} {\bibfield  {journal} {\bibinfo  {journal} {SIAM J. Sci. Comput.}\ }\textbf {\bibinfo {volume} {14}},\ \bibinfo {pages} {1368} (\bibinfo {year} {1993})}\BibitemShut {NoStop}%
\bibitem [{\citenamefont {Greengard}\ and\ \citenamefont {Lee}(2004)}]{greengard2004accelerating}%
  \BibitemOpen
  \bibfield  {author} {\bibinfo {author} {\bibfnamefont {L.}~\bibnamefont {Greengard}}\ and\ \bibinfo {author} {\bibfnamefont {J.-Y.}\ \bibnamefont {Lee}},\ }\bibfield  {title} {\bibinfo {title} {Accelerating the nonuniform fast {F}ourier transform},\ }\href@noop {} {\bibfield  {journal} {\bibinfo  {journal} {SIAM Rev.}\ }\textbf {\bibinfo {volume} {46}},\ \bibinfo {pages} {443} (\bibinfo {year} {2004})}\BibitemShut {NoStop}%
\bibitem [{\citenamefont {Ko}\ and\ \citenamefont {Ong}(2023)}]{ko2023recent}%
  \BibitemOpen
  \bibfield  {author} {\bibinfo {author} {\bibfnamefont {T.~W.}\ \bibnamefont {Ko}}\ and\ \bibinfo {author} {\bibfnamefont {S.~P.}\ \bibnamefont {Ong}},\ }\bibfield  {title} {\bibinfo {title} {Recent advances and outstanding challenges for machine learning interatomic potentials},\ }\href@noop {} {\bibfield  {journal} {\bibinfo  {journal} {Nat. Comput. Sci.}\ }\textbf {\bibinfo {volume} {3}},\ \bibinfo {pages} {998} (\bibinfo {year} {2023})}\BibitemShut {NoStop}%
\bibitem [{sup()}]{supplementary_information}%
  \BibitemOpen
  \href@noop {} {\bibinfo {title} {See {Supplementary Material} for detailed mathematical derivations, numerical quadrature scheme, its error analysis, and numerical validations}},\ \bibinfo {howpublished} {\url{[url]}},\ \bibinfo {note} {includes references [3,\,8,\,10,\,13,\,14,\,19,\,20,\,23,\,24,\,26,\,27,\,30-33,\,37,~38,\,50,\,51]}\BibitemShut {NoStop}%
\bibitem [{\citenamefont {Folland}\ and\ \citenamefont {Sitaram}(1997)}]{folland1997uncertainty}%
  \BibitemOpen
  \bibfield  {author} {\bibinfo {author} {\bibfnamefont {G.~B.}\ \bibnamefont {Folland}}\ and\ \bibinfo {author} {\bibfnamefont {A.}~\bibnamefont {Sitaram}},\ }\bibfield  {title} {\bibinfo {title} {The uncertainty principle: a mathematical survey},\ }\href@noop {} {\bibfield  {journal} {\bibinfo  {journal} {J. Fourier Anal. App.}\ }\textbf {\bibinfo {volume} {3}},\ \bibinfo {pages} {207} (\bibinfo {year} {1997})}\BibitemShut {NoStop}%
\bibitem [{\citenamefont {Faller}\ \emph {et~al.}(2024)\citenamefont {Faller}, \citenamefont {Kaltak},\ and\ \citenamefont {Kresse}}]{faller2024density}%
  \BibitemOpen
  \bibfield  {author} {\bibinfo {author} {\bibfnamefont {C.}~\bibnamefont {Faller}}, \bibinfo {author} {\bibfnamefont {M.}~\bibnamefont {Kaltak}},\ and\ \bibinfo {author} {\bibfnamefont {G.}~\bibnamefont {Kresse}},\ }\bibfield  {title} {\bibinfo {title} {Density-based long-range electrostatic descriptors for machine learning force fields},\ }\href@noop {} {\bibfield  {journal} {\bibinfo  {journal} {J. Chem. Phys.}\ }\textbf {\bibinfo {volume} {161}},\ \bibinfo {pages} {214701} (\bibinfo {year} {2024})}\BibitemShut {NoStop}%
\bibitem [{\citenamefont {Benner}\ \emph {et~al.}(2015)\citenamefont {Benner}, \citenamefont {Gugercin},\ and\ \citenamefont {Willcox}}]{benner2015survey}%
  \BibitemOpen
  \bibfield  {author} {\bibinfo {author} {\bibfnamefont {P.}~\bibnamefont {Benner}}, \bibinfo {author} {\bibfnamefont {S.}~\bibnamefont {Gugercin}},\ and\ \bibinfo {author} {\bibfnamefont {K.}~\bibnamefont {Willcox}},\ }\bibfield  {title} {\bibinfo {title} {A survey of projection-based model reduction methods for parametric dynamical systems},\ }\href@noop {} {\bibfield  {journal} {\bibinfo  {journal} {SIAM Rev.}\ }\textbf {\bibinfo {volume} {57}},\ \bibinfo {pages} {483} (\bibinfo {year} {2015})}\BibitemShut {NoStop}%
\bibitem [{\citenamefont {Thompson}\ \emph {et~al.}(2022)\citenamefont {Thompson}, \citenamefont {Aktulga}, \citenamefont {Berger}, \citenamefont {Bolintineanu}, \citenamefont {Brown}, \citenamefont {Crozier}, \citenamefont {In't~Veld}, \citenamefont {Kohlmeyer}, \citenamefont {Moore}, \citenamefont {Nguyen} \emph {et~al.}}]{thompson2022lammps}%
  \BibitemOpen
  \bibfield  {author} {\bibinfo {author} {\bibfnamefont {A.~P.}\ \bibnamefont {Thompson}}, \bibinfo {author} {\bibfnamefont {H.~M.}\ \bibnamefont {Aktulga}}, \bibinfo {author} {\bibfnamefont {R.}~\bibnamefont {Berger}}, \bibinfo {author} {\bibfnamefont {D.~S.}\ \bibnamefont {Bolintineanu}}, \bibinfo {author} {\bibfnamefont {W.~M.}\ \bibnamefont {Brown}}, \bibinfo {author} {\bibfnamefont {P.~S.}\ \bibnamefont {Crozier}}, \bibinfo {author} {\bibfnamefont {P.~J.}\ \bibnamefont {In't~Veld}}, \bibinfo {author} {\bibfnamefont {A.}~\bibnamefont {Kohlmeyer}}, \bibinfo {author} {\bibfnamefont {S.~G.}\ \bibnamefont {Moore}}, \bibinfo {author} {\bibfnamefont {T.~D.}\ \bibnamefont {Nguyen}}, \emph {et~al.},\ }\bibfield  {title} {\bibinfo {title} {{LAMMPS}-a flexible simulation tool for particle-based materials modeling at the atomic, meso, and continuum scales},\ }\href@noop {} {\bibfield  {journal} {\bibinfo  {journal} {Comput. Phys. Commun.}\ }\textbf {\bibinfo {volume} {271}},\ \bibinfo {pages} {108171} (\bibinfo
  {year} {2022})}\BibitemShut {NoStop}%
\bibitem [{\citenamefont {Burns}\ \emph {et~al.}(2017)\citenamefont {Burns}, \citenamefont {Faver}, \citenamefont {Zheng}, \citenamefont {Marshall}, \citenamefont {Smith}, \citenamefont {Vanommeslaeghe}, \citenamefont {MacKerell}, \citenamefont {Merz},\ and\ \citenamefont {Sherrill}}]{burns2017biofragment}%
  \BibitemOpen
  \bibfield  {author} {\bibinfo {author} {\bibfnamefont {L.~A.}\ \bibnamefont {Burns}}, \bibinfo {author} {\bibfnamefont {J.~C.}\ \bibnamefont {Faver}}, \bibinfo {author} {\bibfnamefont {Z.}~\bibnamefont {Zheng}}, \bibinfo {author} {\bibfnamefont {M.~S.}\ \bibnamefont {Marshall}}, \bibinfo {author} {\bibfnamefont {D.~G.}\ \bibnamefont {Smith}}, \bibinfo {author} {\bibfnamefont {K.}~\bibnamefont {Vanommeslaeghe}}, \bibinfo {author} {\bibfnamefont {A.~D.}\ \bibnamefont {MacKerell}}, \bibinfo {author} {\bibfnamefont {K.~M.}\ \bibnamefont {Merz}},\ and\ \bibinfo {author} {\bibfnamefont {C.~D.}\ \bibnamefont {Sherrill}},\ }\bibfield  {title} {\bibinfo {title} {{The BioFragment Database (BFDb): An open-data platform for computational chemistry analysis of noncovalent interactions}},\ }\href@noop {} {\bibfield  {journal} {\bibinfo  {journal} {J. Chem. Phys.}\ }\textbf {\bibinfo {volume} {147}},\ \bibinfo {pages} {161727} (\bibinfo {year} {2017})}\BibitemShut {NoStop}%
\bibitem [{\citenamefont {Heyd}\ \emph {et~al.}(2003)\citenamefont {Heyd}, \citenamefont {Scuseria},\ and\ \citenamefont {Ernzerhof}}]{heyd2003hybrid}%
  \BibitemOpen
  \bibfield  {author} {\bibinfo {author} {\bibfnamefont {J.}~\bibnamefont {Heyd}}, \bibinfo {author} {\bibfnamefont {G.~E.}\ \bibnamefont {Scuseria}},\ and\ \bibinfo {author} {\bibfnamefont {M.}~\bibnamefont {Ernzerhof}},\ }\bibfield  {title} {\bibinfo {title} {Hybrid functionals based on a screened {C}oulomb potential},\ }\href@noop {} {\bibfield  {journal} {\bibinfo  {journal} {J. Chem. Phys.}\ }\textbf {\bibinfo {volume} {118}},\ \bibinfo {pages} {8207} (\bibinfo {year} {2003})}\BibitemShut {NoStop}%
\bibitem [{\citenamefont {Cheng}(2025)}]{cheng2024latent}%
  \BibitemOpen
  \bibfield  {author} {\bibinfo {author} {\bibfnamefont {B.}~\bibnamefont {Cheng}},\ }\bibfield  {title} {\bibinfo {title} {Latent {E}wald summation for machine learning of long-range interactions},\ }\href@noop {} {\bibfield  {journal} {\bibinfo  {journal} {npj Comput. Mater.}\ }\textbf {\bibinfo {volume} {11}},\ \bibinfo {pages} {80} (\bibinfo {year} {2025})}\BibitemShut {NoStop}%
\bibitem [{\citenamefont {Kim}\ \emph {et~al.}(2024)\citenamefont {Kim}, \citenamefont {King}, \citenamefont {Zhong},\ and\ \citenamefont {Cheng}}]{kim2024learning}%
  \BibitemOpen
  \bibfield  {author} {\bibinfo {author} {\bibfnamefont {D.}~\bibnamefont {Kim}}, \bibinfo {author} {\bibfnamefont {D.~S.}\ \bibnamefont {King}}, \bibinfo {author} {\bibfnamefont {P.}~\bibnamefont {Zhong}},\ and\ \bibinfo {author} {\bibfnamefont {B.}~\bibnamefont {Cheng}},\ }\bibfield  {title} {\bibinfo {title} {Learning charges and long-range interactions from energies and forces},\ }\href@noop {} {\bibfield  {journal} {\bibinfo  {journal} {arXiv:2412.15455}\ } (\bibinfo {year} {2024})}\BibitemShut {NoStop}%
\bibitem [{\citenamefont {Kresse}\ and\ \citenamefont {Furthm{\"u}ller}(1996)}]{kresse1996efficient}%
  \BibitemOpen
  \bibfield  {author} {\bibinfo {author} {\bibfnamefont {G.}~\bibnamefont {Kresse}}\ and\ \bibinfo {author} {\bibfnamefont {J.}~\bibnamefont {Furthm{\"u}ller}},\ }\bibfield  {title} {\bibinfo {title} {Efficient iterative schemes for ab initio total-energy calculations using a plane-wave basis set},\ }\href@noop {} {\bibfield  {journal} {\bibinfo  {journal} {Phys. Rev. B}\ }\textbf {\bibinfo {volume} {54}},\ \bibinfo {pages} {11169} (\bibinfo {year} {1996})}\BibitemShut {NoStop}%
\bibitem [{\citenamefont {Payne}\ \emph {et~al.}(1992)\citenamefont {Payne}, \citenamefont {Teter}, \citenamefont {Allan}, \citenamefont {Arias},\ and\ \citenamefont {Joannopoulos}}]{payne1992iterative}%
  \BibitemOpen
  \bibfield  {author} {\bibinfo {author} {\bibfnamefont {M.~C.}\ \bibnamefont {Payne}}, \bibinfo {author} {\bibfnamefont {M.~P.}\ \bibnamefont {Teter}}, \bibinfo {author} {\bibfnamefont {D.~C.}\ \bibnamefont {Allan}}, \bibinfo {author} {\bibfnamefont {T.}~\bibnamefont {Arias}},\ and\ \bibinfo {author} {\bibfnamefont {a.~J.}\ \bibnamefont {Joannopoulos}},\ }\bibfield  {title} {\bibinfo {title} {Iterative minimization techniques for ab initio total-energy calculations: molecular dynamics and conjugate gradients},\ }\href@noop {} {\bibfield  {journal} {\bibinfo  {journal} {Rev. Mod. Phys.}\ }\textbf {\bibinfo {volume} {64}},\ \bibinfo {pages} {1045} (\bibinfo {year} {1992})}\BibitemShut {NoStop}%
\bibitem [{\citenamefont {Hu}(2022)}]{hu2022symmetry}%
  \BibitemOpen
  \bibfield  {author} {\bibinfo {author} {\bibfnamefont {Z.}~\bibnamefont {Hu}},\ }\bibfield  {title} {\bibinfo {title} {The symmetry-preserving mean field condition for electrostatic correlations in bulk},\ }\href@noop {} {\bibfield  {journal} {\bibinfo  {journal} {J. Chem. Phys.}\ }\textbf {\bibinfo {volume} {156}} (\bibinfo {year} {2022})}\BibitemShut {NoStop}%
\bibitem [{\citenamefont {Cox}(2020)}]{cox2020dielectric}%
  \BibitemOpen
  \bibfield  {author} {\bibinfo {author} {\bibfnamefont {S.~J.}\ \bibnamefont {Cox}},\ }\bibfield  {title} {\bibinfo {title} {Dielectric response with short-ranged electrostatics},\ }\href@noop {} {\bibfield  {journal} {\bibinfo  {journal} {P. Nat. Acad. Sci.}\ }\textbf {\bibinfo {volume} {117}},\ \bibinfo {pages} {19746} (\bibinfo {year} {2020})}\BibitemShut {NoStop}%
\bibitem [{\citenamefont {Schlaich}\ \emph {et~al.}(2016)\citenamefont {Schlaich}, \citenamefont {Knapp},\ and\ \citenamefont {Netz}}]{schlaich2016water}%
  \BibitemOpen
  \bibfield  {author} {\bibinfo {author} {\bibfnamefont {A.}~\bibnamefont {Schlaich}}, \bibinfo {author} {\bibfnamefont {E.~W.}\ \bibnamefont {Knapp}},\ and\ \bibinfo {author} {\bibfnamefont {R.~R.}\ \bibnamefont {Netz}},\ }\bibfield  {title} {\bibinfo {title} {Water dielectric effects in planar confinement},\ }\href@noop {} {\bibfield  {journal} {\bibinfo  {journal} {Phys. Rev. Lett.}\ }\textbf {\bibinfo {volume} {117}},\ \bibinfo {pages} {048001} (\bibinfo {year} {2016})}\BibitemShut {NoStop}%
\bibitem [{\citenamefont {Rinaldi}\ \emph {et~al.}(2025)\citenamefont {Rinaldi}, \citenamefont {Bochkarev}, \citenamefont {Lysogorskiy},\ and\ \citenamefont {Drautz}}]{rinaldi2025charge}%
  \BibitemOpen
  \bibfield  {author} {\bibinfo {author} {\bibfnamefont {M.}~\bibnamefont {Rinaldi}}, \bibinfo {author} {\bibfnamefont {A.}~\bibnamefont {Bochkarev}}, \bibinfo {author} {\bibfnamefont {Y.}~\bibnamefont {Lysogorskiy}},\ and\ \bibinfo {author} {\bibfnamefont {R.}~\bibnamefont {Drautz}},\ }\bibfield  {title} {\bibinfo {title} {Charge-constrained atomic cluster expansion},\ }\href@noop {} {\bibfield  {journal} {\bibinfo  {journal} {Phys. Rev. Mater.}\ }\textbf {\bibinfo {volume} {9}},\ \bibinfo {pages} {033802} (\bibinfo {year} {2025})}\BibitemShut {NoStop}%
\bibitem [{\citenamefont {Eastman}\ \emph {et~al.}(2023)\citenamefont {Eastman}, \citenamefont {Behara}, \citenamefont {Dotson}, \citenamefont {Galvelis}, \citenamefont {Herr}, \citenamefont {Horton}, \citenamefont {Mao}, \citenamefont {Chodera}, \citenamefont {Pritchard}, \citenamefont {Wang} \emph {et~al.}}]{eastman2023spice}%
  \BibitemOpen
  \bibfield  {author} {\bibinfo {author} {\bibfnamefont {P.}~\bibnamefont {Eastman}}, \bibinfo {author} {\bibfnamefont {P.~K.}\ \bibnamefont {Behara}}, \bibinfo {author} {\bibfnamefont {D.~L.}\ \bibnamefont {Dotson}}, \bibinfo {author} {\bibfnamefont {R.}~\bibnamefont {Galvelis}}, \bibinfo {author} {\bibfnamefont {J.~E.}\ \bibnamefont {Herr}}, \bibinfo {author} {\bibfnamefont {J.~T.}\ \bibnamefont {Horton}}, \bibinfo {author} {\bibfnamefont {Y.}~\bibnamefont {Mao}}, \bibinfo {author} {\bibfnamefont {J.~D.}\ \bibnamefont {Chodera}}, \bibinfo {author} {\bibfnamefont {B.~P.}\ \bibnamefont {Pritchard}}, \bibinfo {author} {\bibfnamefont {Y.}~\bibnamefont {Wang}}, \emph {et~al.},\ }\bibfield  {title} {\bibinfo {title} {{SPICE}, a dataset of drug-like molecules and peptides for training machine learning potentials},\ }\href@noop {} {\bibfield  {journal} {\bibinfo  {journal} {Sci. Data}\ }\textbf {\bibinfo {volume} {10}},\ \bibinfo {pages} {11} (\bibinfo {year} {2023})}\BibitemShut {NoStop}%
\bibitem [{\citenamefont {Liang}\ \emph {et~al.}(2021)\citenamefont {Liang}, \citenamefont {Xu},\ and\ \citenamefont {Zhao}}]{liang2021random}%
  \BibitemOpen
  \bibfield  {author} {\bibinfo {author} {\bibfnamefont {J.}~\bibnamefont {Liang}}, \bibinfo {author} {\bibfnamefont {Z.}~\bibnamefont {Xu}},\ and\ \bibinfo {author} {\bibfnamefont {Y.}~\bibnamefont {Zhao}},\ }\bibfield  {title} {\bibinfo {title} {Random-batch list algorithm for short-range molecular dynamics simulations},\ }\href@noop {} {\bibfield  {journal} {\bibinfo  {journal} {J. Chem. Phys.}\ }\textbf {\bibinfo {volume} {155}},\ \bibinfo {pages} {044108} (\bibinfo {year} {2021})}\BibitemShut {NoStop}%
\bibitem [{\citenamefont {Liang}\ \emph {et~al.}(2023)\citenamefont {Liang}, \citenamefont {Xu},\ and\ \citenamefont {Zhou}}]{liang2023random}%
  \BibitemOpen
  \bibfield  {author} {\bibinfo {author} {\bibfnamefont {J.}~\bibnamefont {Liang}}, \bibinfo {author} {\bibfnamefont {Z.}~\bibnamefont {Xu}},\ and\ \bibinfo {author} {\bibfnamefont {Q.}~\bibnamefont {Zhou}},\ }\bibfield  {title} {\bibinfo {title} {{Random batch sum-of-Gaussians method for molecular dynamics simulations of particle systems}},\ }\href@noop {} {\bibfield  {journal} {\bibinfo  {journal} {SIAM J. Sci. Comput.}\ }\textbf {\bibinfo {volume} {45}},\ \bibinfo {pages} {B591} (\bibinfo {year} {2023})}\BibitemShut {NoStop}%
\bibitem [{\citenamefont {Behler}\ and\ \citenamefont {Parrinello}(2007)}]{behler2007generalized}%
  \BibitemOpen
  \bibfield  {author} {\bibinfo {author} {\bibfnamefont {J.}~\bibnamefont {Behler}}\ and\ \bibinfo {author} {\bibfnamefont {M.}~\bibnamefont {Parrinello}},\ }\bibfield  {title} {\bibinfo {title} {Generalized neural-network representation of high-dimensional potential-energy surfaces},\ }\href@noop {} {\bibfield  {journal} {\bibinfo  {journal} {Phys. Rev. Lett.}\ }\textbf {\bibinfo {volume} {98}},\ \bibinfo {pages} {146401} (\bibinfo {year} {2007})}\BibitemShut {NoStop}%
\bibitem [{\citenamefont {Drautz}(2019)}]{drautz2019atomic}%
  \BibitemOpen
  \bibfield  {author} {\bibinfo {author} {\bibfnamefont {R.}~\bibnamefont {Drautz}},\ }\bibfield  {title} {\bibinfo {title} {Atomic cluster expansion for accurate and transferable interatomic potentials},\ }\href@noop {} {\bibfield  {journal} {\bibinfo  {journal} {Phys. Rev. B}\ }\textbf {\bibinfo {volume} {99}},\ \bibinfo {pages} {014104} (\bibinfo {year} {2019})}\BibitemShut {NoStop}%
\bibitem [{\citenamefont {Cheng}(2024)}]{cheng2024cartesian}%
  \BibitemOpen
  \bibfield  {author} {\bibinfo {author} {\bibfnamefont {B.}~\bibnamefont {Cheng}},\ }\bibfield  {title} {\bibinfo {title} {Cartesian atomic cluster expansion for machine learning interatomic potentials},\ }\href@noop {} {\bibfield  {journal} {\bibinfo  {journal} {npj Comput. Mater.}\ }\textbf {\bibinfo {volume} {10}},\ \bibinfo {pages} {157} (\bibinfo {year} {2024})}\BibitemShut {NoStop}%
\bibitem [{\citenamefont {Shapeev}(2016)}]{shapeev2016moment}%
  \BibitemOpen
  \bibfield  {author} {\bibinfo {author} {\bibfnamefont {A.~V.}\ \bibnamefont {Shapeev}},\ }\bibfield  {title} {\bibinfo {title} {Moment tensor potentials: {A} class of systematically improvable interatomic potentials},\ }\href@noop {} {\bibfield  {journal} {\bibinfo  {journal} {Multiscale Model. Simul.}\ }\textbf {\bibinfo {volume} {14}},\ \bibinfo {pages} {1153} (\bibinfo {year} {2016})}\BibitemShut {NoStop}%
\bibitem [{\citenamefont {Rappe}\ and\ \citenamefont {Goddard~III}(1991)}]{rappe1991charge}%
  \BibitemOpen
  \bibfield  {author} {\bibinfo {author} {\bibfnamefont {A.~K.}\ \bibnamefont {Rappe}}\ and\ \bibinfo {author} {\bibfnamefont {W.~A.}\ \bibnamefont {Goddard~III}},\ }\bibfield  {title} {\bibinfo {title} {Charge equilibration for molecular dynamics simulations},\ }\href@noop {} {\bibfield  {journal} {\bibinfo  {journal} {J. Phys. Chem.}\ }\textbf {\bibinfo {volume} {95}},\ \bibinfo {pages} {3358} (\bibinfo {year} {1991})}\BibitemShut {NoStop}%
\bibitem [{\citenamefont {Ji}\ \emph {et~al.}(2025)\citenamefont {Ji}, \citenamefont {Liang},\ and\ \citenamefont {Xu}}]{SOG-Ne}%
  \BibitemOpen
  \bibfield  {author} {\bibinfo {author} {\bibfnamefont {Y.}~\bibnamefont {Ji}}, \bibinfo {author} {\bibfnamefont {J.}~\bibnamefont {Liang}},\ and\ \bibinfo {author} {\bibfnamefont {Z.}~\bibnamefont {Xu}},\ }\href@noop {} {\bibinfo {title} {{ Supplementary data for ``Machine-Learning Interatomic Potentials for Long-Range Systems''}}},\ \bibinfo {howpublished} {\url{https://github.com/DuktigYajie/SOG-Net}} (\bibinfo {year} {2025})\BibitemShut {NoStop}%
\bibitem [{\citenamefont {Darden}\ \emph {et~al.}(1993)\citenamefont {Darden}, \citenamefont {York},\ and\ \citenamefont {Pedersen}}]{Darden1993JCP}%
  \BibitemOpen
  \bibfield  {author} {\bibinfo {author} {\bibfnamefont {T.}~\bibnamefont {Darden}}, \bibinfo {author} {\bibfnamefont {D.}~\bibnamefont {York}},\ and\ \bibinfo {author} {\bibfnamefont {L.}~\bibnamefont {Pedersen}},\ }\bibfield  {title} {\bibinfo {title} {{Particle mesh Ewald: An $N\cdot \log(N)$ method for Ewald sums in large systems}},\ }\href@noop {} {\bibfield  {journal} {\bibinfo  {journal} {J. Chem. Phys.}\ }\textbf {\bibinfo {volume} {98}},\ \bibinfo {pages} {10089} (\bibinfo {year} {1993})}\BibitemShut {NoStop}%
\bibitem [{\citenamefont {Hockney}\ and\ \citenamefont {Eastwood}(1988)}]{Hockney1988Computer}%
  \BibitemOpen
  \bibfield  {author} {\bibinfo {author} {\bibfnamefont {R.~W.}\ \bibnamefont {Hockney}}\ and\ \bibinfo {author} {\bibfnamefont {J.~W.}\ \bibnamefont {Eastwood}},\ }\href@noop {} {\emph {\bibinfo {title} {{Computer Simulation Using Particles}}}}\ (\bibinfo  {publisher} {CRC Press},\ \bibinfo {year} {1988})\BibitemShut {NoStop}%
\bibitem [{\citenamefont {Greengard}\ and\ \citenamefont {Rokhlin}(1987)}]{greengard1987fast}%
  \BibitemOpen
  \bibfield  {author} {\bibinfo {author} {\bibfnamefont {L.}~\bibnamefont {Greengard}}\ and\ \bibinfo {author} {\bibfnamefont {V.}~\bibnamefont {Rokhlin}},\ }\bibfield  {title} {\bibinfo {title} {A fast algorithm for particle simulations},\ }\href@noop {} {\bibfield  {journal} {\bibinfo  {journal} {J. Comput. Phys.}\ }\textbf {\bibinfo {volume} {73}},\ \bibinfo {pages} {325} (\bibinfo {year} {1987})}\BibitemShut {NoStop}%
\bibitem [{\citenamefont {Greengard}(1988)}]{greengard1988}%
  \BibitemOpen
  \bibfield  {author} {\bibinfo {author} {\bibfnamefont {L.}~\bibnamefont {Greengard}},\ }\href@noop {} {\emph {\bibinfo {title} {The rapid evaluation of potential fields in particle systems}}}\ (\bibinfo  {publisher} {MIT Press},\ \bibinfo {year} {1988})\BibitemShut {NoStop}%
\bibitem [{\citenamefont {Jin}\ \emph {et~al.}(2021)\citenamefont {Jin}, \citenamefont {Li}, \citenamefont {Xu},\ and\ \citenamefont {Zhao}}]{jin2021random}%
  \BibitemOpen
  \bibfield  {author} {\bibinfo {author} {\bibfnamefont {S.}~\bibnamefont {Jin}}, \bibinfo {author} {\bibfnamefont {L.}~\bibnamefont {Li}}, \bibinfo {author} {\bibfnamefont {Z.}~\bibnamefont {Xu}},\ and\ \bibinfo {author} {\bibfnamefont {Y.}~\bibnamefont {Zhao}},\ }\bibfield  {title} {\bibinfo {title} {A random batch {Ewald} method for particle systems with {Coulomb} interactions},\ }\href@noop {} {\bibfield  {journal} {\bibinfo  {journal} {SIAM J. Sci. Comput.}\ }\textbf {\bibinfo {volume} {43}},\ \bibinfo {pages} {B937} (\bibinfo {year} {2021})}\BibitemShut {NoStop}%
\bibitem [{\citenamefont {Wolf}\ \emph {et~al.}(1999)\citenamefont {Wolf}, \citenamefont {Keblinski}, \citenamefont {Phillpot},\ and\ \citenamefont {Eggebrecht}}]{wolf1999exact}%
  \BibitemOpen
  \bibfield  {author} {\bibinfo {author} {\bibfnamefont {D.}~\bibnamefont {Wolf}}, \bibinfo {author} {\bibfnamefont {P.}~\bibnamefont {Keblinski}}, \bibinfo {author} {\bibfnamefont {S.}~\bibnamefont {Phillpot}},\ and\ \bibinfo {author} {\bibfnamefont {J.}~\bibnamefont {Eggebrecht}},\ }\bibfield  {title} {\bibinfo {title} {{Exact method for the simulation of Coulombic systems by spherically truncated, pairwise $r^{-1}$ summation}},\ }\href@noop {} {\bibfield  {journal} {\bibinfo  {journal} {J. Chem. Phys.}\ }\textbf {\bibinfo {volume} {110}},\ \bibinfo {pages} {8254} (\bibinfo {year} {1999})}\BibitemShut {NoStop}%
\bibitem [{\citenamefont {Fennell}\ and\ \citenamefont {Gezelter}(2006)}]{fennell2006ewald}%
  \BibitemOpen
  \bibfield  {author} {\bibinfo {author} {\bibfnamefont {C.~J.}\ \bibnamefont {Fennell}}\ and\ \bibinfo {author} {\bibfnamefont {J.~D.}\ \bibnamefont {Gezelter}},\ }\bibfield  {title} {\bibinfo {title} {{Is the Ewald summation still necessary? Pairwise alternatives to the accepted standard for long-range electrostatics}},\ }\href@noop {} {\bibfield  {journal} {\bibinfo  {journal} {J. Chem. Phys.}\ }\textbf {\bibinfo {volume} {124}},\ \bibinfo {pages} {234104} (\bibinfo {year} {2006})}\BibitemShut {NoStop}%
\bibitem [{\citenamefont {Barker}\ and\ \citenamefont {Watts}(1973)}]{barker1973monte}%
  \BibitemOpen
  \bibfield  {author} {\bibinfo {author} {\bibfnamefont {J.~A.}\ \bibnamefont {Barker}}\ and\ \bibinfo {author} {\bibfnamefont {R.~O.}\ \bibnamefont {Watts}},\ }\bibfield  {title} {\bibinfo {title} {{Monte Carlo studies of the dielectric properties of water-like models}},\ }\href@noop {} {\bibfield  {journal} {\bibinfo  {journal} {Mol. Phys.}\ }\textbf {\bibinfo {volume} {26}},\ \bibinfo {pages} {789} (\bibinfo {year} {1973})}\BibitemShut {NoStop}%
\bibitem [{\citenamefont {Barnett}\ \emph {et~al.}(2019)\citenamefont {Barnett}, \citenamefont {Magland},\ and\ \citenamefont {af~Klinteberg}}]{Barnett2019SISC}%
  \BibitemOpen
  \bibfield  {author} {\bibinfo {author} {\bibfnamefont {A.~H.}\ \bibnamefont {Barnett}}, \bibinfo {author} {\bibfnamefont {J.}~\bibnamefont {Magland}},\ and\ \bibinfo {author} {\bibfnamefont {L.}~\bibnamefont {af~Klinteberg}},\ }\bibfield  {title} {\bibinfo {title} {A parallel nonuniform fast {F}ourier transform library based on an ``exponential of semicircle'' kernel},\ }\href@noop {} {\bibfield  {journal} {\bibinfo  {journal} {SIAM J. Sci. Comput.}\ }\textbf {\bibinfo {volume} {41}},\ \bibinfo {pages} {C479} (\bibinfo {year} {2019})}\BibitemShut {NoStop}%
\end{thebibliography}
\end{document}